**Digital Sovereigns: Big Tech's and Nation-State Influence**

A thesis presented to

Norwich University

In partial fulfillment

of the requirements for the degree

Master of Arts in International Relations

Michael Bollerman

November 10,  2024

# Table of Contents

# Abstract

BOLLERMAN, MICHAEL , MASTER OF ARTS IN INTERNATIONAL STUDIES ,

November 10, 2024

<u>Digital Sovereigns: Big Tech's and Nation-State Influence</u>

Technology companies have gained unprecedented power and influence in recent years, resembling quasi-nation-states globally. Corporations like those with trillion-dollar market capitalization are no longer just providers of digital services; they now wield immense economic power, influence global infrastructure, and significantly impact political and social dynamics. This thesis delves into how these corporations have transcended traditional business models, taking on characteristics typically associated with sovereign states. They now enforce regulations, shape public discourse, and influence legal frameworks in various countries. This shift presents unique challenges, including the undermining of democratic governance, exacerbation of economic inequalities, and enabling unregulated data exploitation and privacy violations. The study will examine critical instances of tech companies acting as quasi-governmental bodies and evaluate the risks of unchecked corporate influence in global governance. Ultimately, the thesis aims to propose policy frameworks and regulatory interventions to curb the overreach of tech giants, restoring the balance between democratic institutions and corporate power and ensuring that the digital future aligns with the public good rather than creating Frankenstein-like monsters.

# Dedication

*To technocrats everywhere who cannot live without their devices and wonder how and in what ways the companies that produce these devices might enhance or harm us.*

# 1. Introduction

## 1.1 Opening Statement: Nation-States and the Rise of Tech Giants

Technology has become essential in our lives, providing unlimited access to information, increasing productivity in the workplace, revolutionizing communication, aiding education, unlocking avenues for creativity, and playing a pivotal role in research and development. Using technology wisely and responsibly is essential to ensure its long-term positive benefits.[1]  In the modern era, technology companies have become pivotal actors on the global stage, rivaling traditional nation-states in terms of influence and power. The largest firms—Apple, Broadcom, Meta Platforms (Facebook), Microsoft, NVIDIA, TSMC, X Corp., and Tesla—have transcended their initial roles as service providers, amassing considerable economic, social, and political sway. This transformation raises significant concerns about the emergence of these tech giants as quasi-sovereign entities capable of exerting influence over millions, if not billions, of people across borders. As they shape communication, control the flow of information, and gather unprecedented amounts of data, their unchecked power can profoundly reshape global governance and individual freedoms. The rise of these companies is often compared to the historical emergence of powerful industrial conglomerates in the early 20th century. However, unlike previous corporate giants, today's tech firms operate within a digital realm that transcends traditional geographical and regulatory boundaries. Their platforms host virtual economies, regulate speech, and even develop digital currencies rivaling state-controlled fiat money.   In this light, their power resembles the attributes of nation-states, such as territoriality, authority, and control over vital resources (data and the operation of devices) that beg for the comparison that they should be viewed as sovereign digital entities. These companies control technological infrastructure, such as cloud services, search engines, and social media platforms, and the interdependence of such infrastructure provides them with geopolitical leverage typically reserved for governments. With the ability to shape public discourse, influence elections, and control critical digital infrastructure, these tech firms possess the tools of statecraft. The extent of this influence has led some to caution that, without immediate action in the form of appropriate regulatory frameworks, these firms could subvert traditional governance models, creating new, unregulated centers of power that operate beyond the reach of democratic institutions.

---

[1] Science of computer. 2023. "The Benefits of Computers for Humanity: Transforming Lives in the Digital Age." Science of Computer. August 7, 2023. https://scienceofcomputer.com/the-benefits-of-computers-for-humanity-transforming-lives-in-the-digital-age/.

## 1.2 Thesis Statement: Tech Companies as Quasi-Nation States

Governments are facing challenges in regulating large technological companies, and it's becoming evident that the lack of adequate oversight could result in these companies wielding disproportionate power, essentially functioning as independent states. This paper will explore how these firms have acquired such influence and the potential risks of allowing them to operate without checks. They are addressing the rise of trillion-dollar corporations as de facto nation-states call for regulatory, economic, and collaborative strategies. Governments must bolster antitrust laws, enforce stricter regulations, and work with international organizations to ensure accountability. These companies play a pivotal role in shaping legislation, establishing their standards for privacy and security, and leading technological innovations. Their platforms direct the flow of vast amounts of information and commercial activity, thus impacting global markets and societal dynamics. The role of big technology companies in global governance is like that of traditional nation-states. Extensive technology corporations in the trillion-dollar market capitalization range in the present day challenge state sovereignty by becoming their nation-states of a kind with their economic power and control over information. [2] Furthermore, these corporations hold considerable power in global labor markets and have developed sprawling campuses resembling small cities, providing employees with housing, entertainment, healthcare, and other amenities. They have also created comprehensive digital ecosystems, extending from cloud computing to digital payments, ultimately shaping their economic and social environments. [3]

## 1.3 Context and Background: Digital Sovereignty and the Role of Alphabet (Google), Amazon, Apple, Broadcom, Meta Platforms (Facebook), Microsoft, Netflix, NVIDIA, TSMC, X Corp., Tesla, etc.

Sovereignty is traditionally understood as the complete authority of a state over its territory and population,. [4] In the 21st century, as the internet and digital technologies reshaped global interactions, traditional concepts of territorial sovereignty have confronted new challenges. Digital spaces defy physical borders; data flows seamlessly across countries, transcending the reach of any single jurisdiction. This emergence of a transnational digital ecosystem gave rise to what scholars now call digital sovereignty, which seeks to reclaim control over data, digital infrastructure, and the broader cyberspace within a country's domain.

---

[2] Mukul, Jahnavi. "Reimagining Big Tech as Nation States." (2023).
[3] Chesbrough, Henry. "The logic of open innovation: managing intellectual property." California Management Review 45, no. 3 (2003): 33-58.
[4] Nicholls, Sophie. "Sovereignty and Government in Jean Bodin's Six Livres de la République (1576)." Journal of the History of Ideas 80, no. 1 (2019): 47-66. https://dx.doi.org/10.1353/jhi.2019.0002.

Digital sovereignty entails a state's control over its digital ecosystem—namely, data governance, cybersecurity, and internet infrastructure—reflecting a nation's ability to ensure that its digital environment operates independently of foreign influence or interference[5]. This transition toward digital sovereignty marks a shift in power dynamics, no longer constrained by geography but increasingly defined by technological prowess and control over digital resources. Digital sovereignty is increasingly recognized as essential to the future of state power, signifying a shift from traditional territorial authority to control over digital resources. States are becoming more aware that managing digital assets equates to power, prompting efforts to establish cyber boundaries similar to physical borders. As new policies emerge to address issues like data localization, encryption, and the independence of digital infrastructure, digital sovereignty not only represents a modern adaptation of classical sovereignty but also becomes a crucial battleground in the current power struggle between governments and technology giants. This concept of digital sovereignty introduces a novel form of global competition that focuses not on land but on data and control of cyberspace. In today's world, the virtual landscape is as vital as physical territory, leading nations to develop policies that protect their digital futures from foreign nations and corporate influence. Consequently, digital sovereignty is a key issue in the ongoing evolution of global power dynamics, redefining what sovereignty means in the digital age.[6]

---

[5] Maurer, Tim. "Cyber norm emergence at the United Nations." An Analysis of the UN's Activities Regarding Cyber-security. Cambridge, MA: Belfer Center for Science and International Affairs (2011).
[6] DeNardis, Laura. The Global War for Internet Governance. Yale University Press, 2014. See also: Chander, Anupam. The electronic silk road: how the web binds the world together in commerce. Yale University Press, 2013.

With their immense influence, resources, and capabilities, technology companies are increasingly being compared to city-states. Large corporations like Alphabet (Google), Amazon, Apple, Broadcom, Meta Platforms (Facebook), Microsoft, NVIDIA, TSMC, X Corp., and Tesla possess vast economic power, with trillions of dollars in market capitalization and significant global influence that exert power like nation-states, including control over information, monetary influence, and impact on sovereignty.  Recent studies compare the market capitalization of multiple tech giants with countries' annual Gross Domestic Product (GDP) to help better understand the comparison of numbers.[7]  Valued at about three trillion, the Cupertino company is more affluent than 96% of the world. Only seven countries currently outrank the maker of the iPhone financially. Turning an eye to Microsoft, coming at a lesser approximate two trillion valuation, the company still generates enough money to put it on a par with all of Canada. Overall, only nine countries are worth more money than the developer of Windows.   Despite providing a service we likely use nearly daily, Amazon is valued at almost two trillion. This means the company makes more money than 92% of the world.  Alphabet, Google's parent company, is valued at over one trillion and is ahead of all but 12 countries. Companies that don't quite reach the trillion-dollar mark, including Tencent, Tesla, Facebook, and Alibaba, would also be among the wealthiest countries in the world by GDP. Only the United States, China, Japan, Germany, India, the United Kingdom, and France managed to outrank the tech giants.[8]  The environmental impact of these tech companies is profound.  Amazon Web Services recently paid $650 million for another data center to add to its extensive infrastructure. The tech giant is also

---

[7] "Companies Ranked by Market Cap - CompaniesMarketCap.com," n.d., https://companiesmarketcap.com/.
[8] Cohen, Jason. 2021. "World's Most Valuable Tech Companies Are Richer Than Most Countries." PCMAG, April 2, 2021. https://www.pcmag.com/news/worlds-most-valuable-tech-companies-are-richer-than-most-countries.

securing a steady nuclear energy supply to power its operations. The Susquehanna Steam Electric Station outside of Berwick, Pennsylvania, which generates 2.5 gigawatts of nuclear power, has directly powered the data center since it opened in 2023. After striking the deal, Amazon sought to modify the terms of its original agreement to purchase 180 megawatts of additional power directly from the nuclear plant. While Susquehanna agreed to the sale, third parties have challenged the deal, leading to a regulatory dispute that will likely set a precedent for data centers, cryptocurrency mining operations, and other computing facilities with high demands for clean electricity. [9] In a similar story, Constellation Energy plans to reopen Unit 1 of the Three Mile Island nuclear plant in Pennsylvania to provide power to Microsoft under a 20-year agreement. The facility is set to come online in 2028 and will be renamed the Crane Clean Energy Center. This initiative reflects the growing demand for carbon-free power and the resurgence of nuclear energy. The US government is offering subsidies to extend the lives of aging nuclear plants in response to the increasing electricity demand driven by emerging technologies such as AI and electric vehicles. [10] The power consumed, for example, In AI, is enormous. AI uses model training that involves feeding it vast amounts of data and adjusting its parameters to minimize errors. This process requires significant computational power, especially for deep learning models with millions or billions of parameters. [11] AI inference, on the other hand, is the process of using a trained model to make predictions or classifications on new data.

---

[9] Moseman, Andrew. 2024. "Amazon Vies for Nuclear-Powered Data Center." IEEE Spectrum, September 26, 2024. https://spectrum.ieee.org/amazon-data-center-nuclear-power.

[10] Bellan, Rebecca. 2024. "Microsoft Taps Three Mile Island Nuclear Plant to Power AI." TechCrunch, September 20, 2024. https://techcrunch.com/2024/09/20/microsoft-taps-three-mile-island-nuclear-plant-to-power-ai/.

[11] Chen, Michael. 2023. "What Is AI Model Training & Why Is It Important?" December 6, 2023. https://www.oracle.com/artificial-intelligence/ai-model-training/.

While inference requires less computational power than training, it can still be significant depending on the application. [12] These factors influence tech companies to own or restart nuclear power plants with autonomy akin to traditional city-states, significantly shaping economic, social, political, and environmental landscapes. These tech giants not only lead in technological innovations but also dominate, acting as the architects of the digital landscape and dictating the terms for vast amounts of information flow and commercial activity, setting the pace in fields such as AI and autonomous vehicles. These organizations' global operations and cultural impacts are far-reaching, with decisions influencing international markets and societies. Allegations of monopolistic practices, including the unequal distribution of economic benefits, have overshadowed their impressive technological advancements. This underscores the city-state-like behavior of big technology companies and their ability to shape economic, social, and political landscapes. As reported by The Washington Post, more than two million people in the Philippines perform "crowd work" as part of AI's vast underbelly, annually tagging data to ensure language models like ChatGPT function properly. These workers earn low wages, often far below the minimum wage in the Philippines.[13] The rapid growth of corporations, particularly those with market capitalizations near a trillion dollars, has sparked significant debate about their role and impact on global economic and political landscapes. Companies like Apple, Amazon, and Microsoft have amassed resources and influence that surpass those of many sovereign nations. This concentration of economic power raises critical questions about its implications for governance structures, regulatory frameworks, and socioeconomic stability. Despite these corporations' significant contributions, including innovation, job creation, and economic growth, their expansive reach poses challenges. These include regulatory evasion, tax avoidance, labor exploitation, and the potential to undermine democratic processes. Their ability to influence global markets and policies highlights a shift in power dynamics that traditional nation-states may find challenging to manage. They possess financial means, technological infrastructure, and lobbying abilities, allowing them to operate independently of conventional state regulations and borders. This poses risks in regulation, accountability, and global equity. Addressing these issues will require proactive collaboration, new legal and ethical frameworks, and a focus on justice, equity, and sustainability.

---

[12] Martineau, Kim. 2024. "What Is AI Inferencing?" IBM Research. June 27, 2024. https://research.ibm.com/blog/AI-inference-explained.
[13] Tan, Rebecca, and Regine Cabato. 2023. "Behind the AI Boom, an Army of Overseas Workers in 'Digital Sweatshops.'" Washington Post, August 29, 2023. https://www.washingtonpost.com/world/2023/08/28/scale-ai-remotasks-philippines-artificial-intelligence/.

Corporations like Alibaba and Tencent continue to rise as major global players, challenging American tech dominance. However, the need for more transparency in markets like China raises additional concerns. Governments worldwide must address these challenges with a renewed focus on antitrust enforcement, corporate responsibility, and sustainability. The age of mega-corporations calls for new strategies to ensure these powerful entities contribute positively to global society and adhere to ethical standards. Historical cases like the Sherman Antitrust Act of 1890 reveal that concentrated corporate power has long challenged governance structures.[14] Furthermore, companies like Standard Oil and AT&T, which dominated their respective industries, eventually broke up to restore competitive markets[15]. This historical perspective highlights the cyclical nature of corporate growth and regulation, illustrating the ongoing tension between economic innovation and the need for oversight. As today's tech giants wield power reminiscent of past monopolies, they must be held accountable through a combination of historical precedent and modern regulation.

---

[14] Dalrymple, William. The anarchy: The East India Company, corporate violence, and the pillage of an empire. Bloomsbury Publishing USA, 2019.
[15] May, James. "Antitrust practice and procedure in the formative era: The constitutional and conceptual reach of state antitrust law, 1880-1918." U. pa. l. Rev. 135 (1986): 495.

## 1.4 Problem Statement: Corporate Power and Sovereignty in the Digital Age

The rapid growth of corporations, especially those with market capitalizations close to a trillion dollars, has ignited significant debate about their role and impact on the global economic and political landscapes. Companies like Apple, Amazon, and Microsoft have accumulated resources and influence that exceed those of many sovereign nations. This concentration of economic power raises critical questions about its implications for governance structures, regulatory frameworks, and socioeconomic stability. The analogy between Mary Shelley's Frankenstein and today's big technology companies can be understood through several thematic connections. In Frankenstein, Victor Frankenstein creates a being through scientific experimentation, raising ethical concerns about the unintended consequences of pursuing knowledge without responsibility. Similarly, contemporary technology companies, particularly those involved in AI, biotechnology, and social platforms, are often compared to modern-day Frankenstein because they create potent technologies without fully understanding or controlling the societal impacts. Victor Frankenstein abandons his creation, which then wreaks havoc. This reflects concerns about tech companies making products like AI or social media platforms with unforeseen negative consequences, such as privacy violations, misinformation, or job displacement due to automation. Many argue that these companies do not take sufficient responsibility for the full range of societal impacts their technologies generate, highlighting the urgent need for accountability. [16] Like Frankenstein, these tech companies have built something powerful without fully understanding or controlling its outcomes.

Frankenstein's pursuit of scientific glory without considering the ethical ramifications mirrors how tech companies prioritize innovation and growth over societal concerns. For example, the aggressive data collection practices of companies like Facebook and Google have raised concerns about user privacy and manipulation, highlighting how surveillance capitalism undermines human autonomy in ways that echo Frankenstein's lack of foresight about his creation's behavior. [17] In Frankenstein, Victor's unchecked control over life and death is a metaphor for the concentrated power of large tech companies today, which dominate global communications and economies whereby companies such as Amazon, Google, Meta, TikTok, and Twitter/X, to name a few wield disproportionate influence over the information people consume, creating monopolistic structures that diminish individual autonomy. This monopolistic control is akin to Frankenstein's creation gaining strength beyond Victor's control, symbolizing the risks of concentrating too much power in the hands of a few. The themes in Frankenstein around creation, ethical responsibility, and control provide a powerful metaphor for understanding the modern-day challenges posed by large technology companies. At the end of Frankenstein (spoiler alert), Victor Frankenstein dies wishing that he could destroy the monster he created.[18] Bringing ultimate destruction to these companies will not be feasible, even if that is warranted. Solutions to these arguable, at times, problematic monstrosities can be sought. To discern a viable solution, if there is one, one must look and research history for solutions. To quote William Shakespeare from The Tempest, "Whereof what's past is prologue, what to come In yours and my discharge."[19] The technologies are new, but the methodologies and motivations of the companies are similar to those used to gain monetary value, status, and power. However, at the same time, despite this tension between governance and digital technology, a consensus seems to emerge that the current predominant governance approach – a combination of the multi-stakeholder approach as practiced in international institutions with self-regulatory frameworks at the national level – was not practical. Scholars have identified a regulatory deficit that needs to be addressed, whereby self-regulatory laws and regulations lag behind technology companies' operations.[20]

---

[16] Bostrom, N. "Superintelligence: Paths, Dangers, Strategies. Oxford." (2014).
[17] Zuboff, S. (2019). The age of surveillance capitalism: The fight for a human future at the new frontier of power (p. 377). PublicAffairs.
[18] Shelly, Mary. Frankenstein. London: Penguin, 2012.

There are essential distinctions between tech corporations and sovereign states that prevent these companies from truly holding sovereign status. Unlike nation-states, tech companies do not have territorial control, a defining feature of sovereignty. A nation-state has the authority to govern within defined geographic borders, enforce laws, and conduct defense. In contrast, companies like Google or Meta operate across various jurisdictions but do not possess physical control over any territory. While they can influence user behavior through platform policies, they cannot enforce these rules like a sovereign entity would within its borders. Tech companies depend on nation-states for fundamental operational and legal support, such as access to infrastructure, cybersecurity protections, and intellectual property laws. These state-provided resources are essential for these companies to effectively function and enforce their policies. This interdependence starkly distinguishes them from self-sufficient sovereign entities in terms of infrastructure and security. Nation-states monopolize the legitimate use of physical force, a hallmark of sovereignty. In contrast, tech companies cannot exert coercive power in the traditional sense; they can only enforce platform-specific policies, and users can choose to leave the platform if they disagree. In cases of policy breaches, these companies rely on state legal systems to take further action, highlighting their subordinate relationship with state authority and the limits of their power.[21]

---

[19] Ajeyaseelan. 2022. "William Shakespeare (1564–1616) - Collection at Bartleby.com." Collection at Bartleby.Com. November 24, 2022. https://www.bartleby.com/lit-hub/respectfully-quoted/william-shakespeare-15641616-152/.
[20] Bayer, Judit, Bernd Holznagel, Päivi Korpisaari, and Lorna Woods. Perspectives on Platform Regulation: Concepts and Models of Social Media Governance Across the Globe (Volume 1). Nomos Verlagsgesellschaft mbH & Co. KG, 2021.

---

[21] Gillespie, Tarleton. Custodians of the Internet: Platforms, content moderation, and the hidden decisions that shape social media. Yale University Press, 2018.

Although tech companies wield significant economic power and influence global markets, their power is fundamentally economic rather than political. They lack political authority or recognition as independent entities in international relations. This absence of political sovereignty diminishes their influence since they cannot enter into treaties, claim rights under international law, or formally represent the interests of their customers. Tech companies are also accountable to the laws of the states in which they operate. While they may lobby for changes or challenge these laws, they cannot independently dictate or alter them.[22]

## 1.5 Research Objectives: Exploring the Extent of Influence and Implications for Global Governance

The research objective of this thesis is to explore the extent of influence that various global actors exert on international governance structures, and to assess the broader implications of this influence on global governance. This research aims to examine how international institutions, nation-states, multinational corporations contribute to shaping policies and decisions at the global level. Furthermore, it will analyze the power dynamics between these actors and investigate the mechanisms through which they assert their influence.

By focusing on key players within the global governance system, the research will provide a comprehensive analysis of their roles and impact. For instance, research will explore how powerful technology companies influence global governance.  Multinational corporations will also be examined for their growing role in shaping regulations, policies, and transnational advocacy efforts.[23]  A central objective of this research is to understand the broader implications of these influences, especially in terms of human rights and economic governance. As the influence of non-state actors grows, the research will investigate whether this shift enhances or diminishes the legitimacy and accountability of current global governance institutions.[24]The study will also examine how different historic and new governmental actions may act, adapt, or transform under the weight of these influences. By offering both theoretical and empirical insights, this research seeks to contribute to the ongoing debate on the effectiveness and future trajectory of global governance. The findings will provide valuable recommendations for policymakers, international institutions, and civil society actors in navigating the increasingly complex global landscape.

---

[22] Rahman, K. Sabeel, and Kathleen Thelen. "The rise of the platform business model and the transformation of twenty-first-century capitalism." Politics & society 47, no. 2 (2019): 177-204.

The day's news has it that research is available on items such as the ongoing legal battle with Google, as in the United States. Department of Justice considers potential remedies, which is a significant development in the tech industry. The possibility of Google being forced to change its business practices, such as controlling distribution channels and default search engine agreements, could reshape the digital landscape, considering asking a federal judge to break up Google after its ubiquitous search engine was declared an illegal monopoly. The filing is the first step in a monthslong legal process to develop remedies that could reshape a company that's long been synonymous with online search. Google has been facing intensifying regulatory pressure on both sides of the Atlantic. European Union antitrust enforcers also suggested that breaking up the company is the only way to satisfy competition concerns about its digital ad business.[25] Understanding the past is not just about reading today's news; it is a crucial research objective. By delving into the history of corporate power, digital sovereignty, global influence, and tech companies, we can better understand the current issues and their implications.

[23] Scholte, Jan Aart. Globalization: A critical introduction. Bloomsbury Publishing, 2017.
[24] Dingwerth, Klaus, and Jessica F. Green. "Transnationalism." In Research handbook on climate governance, pp. 153-163. Edward Elgar Publishing, 2015.

## 1.6 Overview of Methodology: Economic, Political, Social, and Technological Lenses

The methodology adopts a multi-dimensional approach, analyzing the subject through economic, political, social, and technological lenses to comprehensively explore the influence of Big Tech on nation-states. This interdisciplinary approach allows for a nuanced understanding of how technology companies are reshaping traditional concepts of sovereignty and governance. By combining these four lenses—economic, political, social, and technological—the thesis provides a holistic understanding of how Big Tech companies have transcended their original business domains to act as global powers with influence comparable to that of nation-states. This methodological approach allows for an integrated analysis of how these firms shape global governance, economies, and societies. The economic analysis focuses on the unprecedented financial power of Big Tech companies such as Amazon, Apple, Broadcom, Meta Platforms (Facebook, Instagram, WhatsApp), Microsoft, Netflix, NVIDIA, TSMC, X Corp., Tesla, etc. These companies wield vast economic resources, often surpassing the gross domestic product of many or most countries. The methodology examines how this financial power grants these companies leverage over governments, particularly in tax policy, labor markets, and antitrust regulations. The concept of "surveillance capitalism," where the collection and monetization of personal data by these firms represents a new form of economic domination that is largely unregulated. This study explores how economic dominance allows Big Tech to operate with quasi-sovereign authority, influencing global supply chains and capital flows while largely eluding traditional regulatory frameworks.[26]

---

[25] Chen, Shawn. 2024. "Google Was Declared a Monopoly. The US Suggests Breaking the Company up Among Other Fixes | AP News." AP News. October 9, 2024. https://apnews.com/article/google-search-antitrust-case-59114d8bf1dc4c8453c08acaa4051f14.

From a political perspective, the thesis examines how Big Tech has become a crucial factor in shaping policy and influencing state power. These companies play a significant role in the governance of digital spaces and affect national security policies, democratic processes, and public discourse. The methodology draws from the notion that digital corporations have amassed political clout, often acting in ways that bypass national governments through their ability to regulate online spaces and control critical infrastructure such as cloud computing services and communication networks. Political lobbying efforts and the increasing frequency of public-private partnerships, where companies like Microsoft collaborate with national governments on cybersecurity initiatives.[27] However, tensions exist between nation-states and big tech companies, particularly when firms resist governmental demands for data or adhere to regulatory frameworks that transcend national boundaries.[28] One illustration was when, on December 2, 2016, in San Bernardino, California, Syed Rizwan Farook and Tashfeen Malik, the shooters who attacked Farook's office holiday party, killing 14 and wounding 22, in what officials believe was perhaps at the time an act of terrorism. Farook and Malik were killed in a shootout with police shortly after, and investigators were trying to piece together evidence to pinpoint exactly what motivated the couple and how their plan went undetected.[29] Needing access to the Apple iPhone, investigators required Apple's assistance to break the encryption or unlock the phone. Still, when requested to assist, Apple responded that the order was an "overreach by the U.S. government." [30] The FBI later accessed the phone without Apple's assistance.[31]

---

[26] Zuboff, Shoshana. "The age of surveillance capitalism." In Social theory re-wired, Routledge, 2023.
[27] Culpepper, Pepper D., and Kathleen Thelen. "Are we all Amazon primed? Consumers and the politics of platform power." Comparative Political Studies 53, no. 2 (2020): 288-318.
[28] Bratton, Benjamin H. The stack: On software and sovereignty. MIT press, 2016.

Social implications are critical to this methodology, examining how Big Tech impacts societal norms, values, and behaviors globally. Platforms like Facebook, TikTok, and YouTube shape public opinion, propagate disinformation and influence electoral outcomes. [32] Additionally, the methodology considers how these companies create digital ecosystems that foster community but also exacerbate societal divides and amplify polarization. Issues arise, such as digital rights, online freedom of expression and how Big Tech has increasingly assumed roles traditionally filled by public institutions, such as media companies or public broadcasters. The social analysis delves into the ethical concerns surrounding content moderation, privacy, and the role of algorithms in reinforcing societal biases and inequalities via recommendation engines. The technological lens evaluates how advancements in artificial intelligence (AI), data analytics, and cloud computing enable Big Tech to operate as quasi-sovereign entities. The methodology examines how these technological capabilities give companies unparalleled power to collect, process, and utilize data, allowing them to influence markets and governments.[33] The shift in power dynamics emphasizes how control over data has become a primary form of power in the 21st century. This portion of the thesis analyzes how Big Tech companies set technological development and regulation standards, often outpacing governments' ability to keep up with innovation. It also examines how these firms utilize technological dominance to influence international regulatory bodies and shape global technological norms, impacting cybersecurity, privacy laws, and intellectual property rights.

---

[29] Los Angeles Times. 2015. "Everything We Know About the San Bernardino Terror Attack Investigation so Far - Los Angeles Times," December 14, 2015. https://www.latimes.com/local/california/la-me-san-bernardino-shooting-terror-investigation-htmlstory.html.
[30] Queally, James, and Brian Bennett. 2016. "Apple Opposes Order to Help FBI Unlock Phone Belonging to San Bernardino Shooter - Los Angeles Times." Los Angeles Times, February 19, 2016.
https://www.latimes.com/local/lanow/la-me-ln-fbi-apple-san-bernardino-phone-20160216-story.html.
[31] Clark, Mitchell. 2021. "Here's How the FBI Managed to Get Into the San Bernardino Shooter's iPhone." *The Verge*, April 14, 2021. https://www.theverge.com/2021/4/14/22383957/fbi-san-bernadino-iphone-hack-shooting-investigation.

[32] Vaidhyanathan, Siva. Antisocial media: How Facebook disconnects us and undermines democracy. Oxford University Press, 2018.

### 1.7 Structure of the Thesis

Exploring the evolving role of major technology corporations in global governance, political economies, and their increasing influence over nation-states. The thesis narrative is structured around several critical sections built upon one another to examine the power dynamics between Big Tech and traditional nation-states. The concept of "digital sovereignty," a term that refers to the unprecedented level of control that Big Tech companies have gained in global affairs, outlines the historical context of their rise, starting from the late 20th century when technological advancements and the digital revolution gave birth to these global giants. These companies have amassed vast economic power and influence over digital infrastructure and communication networks by leveraging network effects, data economies, and platform dominance. [34] Conventional theories of political power, sovereignty, and governance must evolve to account for the influence of non-state actors like Big Tech. Drawing on political,

---

[33] Harari, Yuval Noah. *21 Lessons for the 21st Century:'Truly mind-expanding... Ultra-topical'Guardian*. Random House, 2018.
[34] Zuboff, Shoshana. "The age of surveillance capitalism." In Social theory re-wired, pp. 203-213. Routledge, 2023.

economic, and international relations theories both now and in history, will examine concepts

such as soft power which is he ability to influence the behavior of others to get the outcome one

wants"[35] and a form of digital colonialism or appropriating myriad aspects of human life as the

raw material for capitalism[36] of people and their day to day activities to frame the argument that

Big Tech operates with a level of autonomy and power that rivals and, in some cases surpass,

that of nation-states. The economic aspects of Big Tech's rise to power examine how companies have built extensive monopolies and oligopolies through mergers, acquisitions, and innovation.

Companies creating ecosystems that make it difficult for new competitors to enter the markets show how this economic dominance allows Big Tech to influence policy, labor markets, and even taxation in ways that blur the lines between corporate interests and state regulation.

Another illustration of how this story or thesis unfolds while writing is that Meta is in the United

States Federal Trade Commission (FTC) for its possible monopolistic practices. Meanwhile,

Amazon and Microsoft are buying nuclear power plants to help power artificial intelligence (AI)

ambitions.

Big Tech firms have assumed quasi-governmental roles in global politics. Their control over vast amounts of personal data and their ability to shape public discourse and information ecosystems gives them political power to shape domestic and international policies. Google,

Meta and TikTok, and other through their control of social media and search algorithms, have influenced election outcomes and political movements across the globe[37]. Introducing or maybe reinforcing the concept of "digital sovereignty" refers to the unprecedented control that Big Tech companies have gained in global affairs. It outlines the historical context of their rise, starting from the late 20th century when technological advancements and the digital revolution gave birth to these global giants. These companies have amassed vast economic power and influence over digital infrastructure and communication networks by leveraging network effects, data economies, and platform dominance. [38] These companies have built extensive monopolies and oligopolies through mergers, acquisitions, and innovation. As Big Tech's influence has grown, nation-states have attempted to regulate their activities, often with mixed success.

---

[35] Nye, Joseph S. *Soft power: The means to success in world politics*. Public affairs, 2004.
[36] Couldry, Nick, and Ulises A. Mejias. "The costs of connection: How data are colonizing human life and appropriating it for capitalism." (2020): e6-e6.
[37] Vaidhyanathan, Siva. *Antisocial media: How Facebook disconnects us and undermines democracy*. Oxford University Press, 2018.

Speculating on the future relationship between Big Tech and nation-states, there are two potential scenarios: one in which nation-states regain control through stronger regulations and another where Big Tech companies continue to expand their influence, potentially leading to a new global governance model where these corporations hold reins of power. Engaging with contemporary debates about the ethical responsibilities of Big Tech and whether they should be held accountable to global standards akin to nation-states, particularly in areas like human rights and environmental sustainability, especially now that tech companies are using their own nuclear power plants.[39]

## 2. Methodology

### 2.1 Research Design: Qualitative and Quantitative Analysis of Case Studies

There are end amounts of Annual Reports (Form 10-K), Quarterly Reports (Form 10-Q)

---

[38] Zuboff, S. (2019). The Age of Surveillance Capitalism: The Fight for a Human Future at the New Frontier of Power. PublicAffairs.

[39] Gurumurthy, Anita, Deepti Bharthur, Nandini Chami, Jai Vipra, and Ira Anjali Anwar. "Platform planet: Development in the intelligence economy." *Available at SSRN 3872499* (2019).

Insider Trading Filings (Form 4). Proxy Statements (Form DEF 14A). Registration Statements (Form S-1) are available from the SEC.  The best summation of this appears compiled for lobbying is at opensecret.org.

## 2.2 Data Collection: Corporate Reports, Legal Cases, Expert Interviews

As people are constantly using devices, the opportunity for Big Tech to track people has grown.  It started for monetary gain and later became helpful for law enforcement.  Many citizens realize that having devices and a connection to the Internet is a concern because of the lack of privacy caused by mass data collection.  The Telecommunications Act of 1996 was a significant revamp of U.S. telecommunications law. Its goal was to promote competition and reduce regulatory barriers. The act allowed for the deregulation of the broadcasting and telecommunications markets, which enabled more companies to enter the industry. It laid the groundwork for expanding the Internet by promoting access and development. The act was designed to foster innovation, reduce consumer costs, and increase service availability. In 2001, the United States passed The Patriot Act, a shorthand name for The Uniting and Strengthening America by Providing Appropriate Tools Required to Intercept and Obstruct Terrorism Act. The Act was passed shortly after the terrorist attacks on September 11, 2001, and, amongst other things, enhanced law enforcement's surveillance capabilities, including foreign and domestic phone, wire, and computer tapping. The Patriot Act expired in 2015[40]Although it started with arguably good intentions, it eventually became a privacy concern.

In China, the Cybersecurity Law, which came into force in 2017, covers various aspects of network security and has laid the foundation for a comprehensive cybersecurity regulatory regime. This regime, among other things, requires critical infrastructure companies to store data within the People's Republic of China and make this data accessible to intelligence services as needed. The law also allows the Chinese Communist Party (CCP) to compel firms to install backdoors in equipment or software, creating a system of incentives/penalties for compliance.[41] This law sets the stage for a mass surveillance state.  As Chinese coders, influenced by the CCP, inject backdoors, the products from China, which are ubiquitous in part due to low cost and inexpensive labor rates, lead to concern of a worldwide surveillance state. In 2018, the European Union implemented the General Data Protection Regulation (GDPR), one of the world's most stringent data protection laws. The regulation aims to curb tech companies' power over personal data.[42]

---

[40] "Patriot Act." n.d. LII / Legal Information Institute. https://www.law.cornell.edu/wex/patriot_act.

      These companies possess financial and economic power comparable to nation-states, driven by their vast revenues, market capitalization, and influence over global markets. Alphabet is one of the largest corporations in the world, with significant financial operations and global revenue streams from advertising, cloud services, and hardware. Its control over search, ads, and cloud services across continents demonstrates its economic sovereignty. As one of the largest retail and cloud infrastructure providers globally, Amazon controls massive supply chains, logistics, and cloud services. It operates on a scale similar to large economies, employing millions globally. Amazon's ability to influence local governments for favorable tax treatments and its logistic network give it sovereignty-like influence. Apple's supply chains, its power over

---

[41] U.S. Department of Homeland Security. 2020. "Data Security Business Advisory." Data Security Business Advisory. https://www.dhs.gov/sites/default/files/publications/20_1222_data-security-business-advisory.pdf.
[42] "Regulation - 2016/679 - EN - GDPR - EUR-Lex." 2016  https://eur-lex.europa.eu/legal-content/EN/TXT/?uri=CELEX%3A32016R0679.

the global electronics market, and its economic impact on consumers and governments place it in a unique position. It exerts authority through its walled-garden ecosystem of devices and services, much like a nation controls its domestic economy.

These gigantic firms' corporate governance and executive power are centralized, similar to how nation-states are governed. CEOs and board directors act like heads of state and cabinets in many ways. Meta Platforms (Facebook): Mark Zuckerberg's control of Meta, with a dual-class share structure that grants him significant voting power, mimics the centralized control of an authoritarian nation-state leader. Meta governs its ecosystem of users and data in ways that transcend traditional state boundaries, including setting its own "laws" or community standards. With a global customer base and vast intellectual property assets (software, cloud, AI), Microsoft enforces its own rules over its products and services, much like a government establishes and enforces laws within its territory. TSMC (Taiwan Semiconductor Manufacturing Company): As the world's largest semiconductor manufacturer, TSMC wields power over the global technology supply chain. Its critical role in producing chips for devices worldwide gives it geopolitical power like that of a nation with essential natural resources.

These corporations engage in diplomacy and influence global policy-making, often independent of their home countries, just like sovereign states. Alphabet (Google): Alphabet is in policy debates over privacy, AI ethics, and antitrust, frequently lobbying and negotiating with governments worldwide, often on terms similar to nation-to-nation negotiations. Alphabet's influence on global information flow mirrors a nation-state's ability to control media or cultural content. Through its involvement in labor practices, international trade, and data policy, Amazon frequently shapes laws and regulations by lobbying and engaging in negotiations, particularly regarding cloud computing services (AWS), which it provides to governments and private sectors globally. Apple's engagement in data privacy advocacy, taxation policies, and supply chain labor standards highlights how it acts as a negotiator with governments worldwide. Apple's decisions, such as where to manufacture devices, affect employment and economic policies in several countries.

These companies control physical, digital, and market territories like a nation-state controls its borders, infrastructure, and population. Broadcom is a semiconductor company critical to global technology infrastructure. Broadcom's control of hardware and chip supplies affects industries worldwide. Its decisions on mergers and acquisitions, such as attempting to acquire Qualcomm, have geopolitical implications, similar to how a nation's territorial expansion affects global politics. Microsoft controls essential digital infrastructure with its cloud infrastructure (Azure), office software, and operating systems. Governments, enterprises, and individuals worldwide depend on its "territory" of services, much like citizens rely on public services in a nation-state. A recent outage with Microsoft's partner CrowdStrike has shown how dependent we are on Microsoft if there is a failure. NVIDIA dominates the GPU (graphics processing unit) market, especially in sectors like AI, gaming, and data centers, where it controls critical technological advancements. Its influence on AI technology gives it leverage akin to a nation's control over strategic industries like energy or defense. These corporations possess vast amounts of data and control over digital ecosystems, akin to how states control resources and national defense. Meta's ownership of user data across its platforms (Facebook, Instagram, WhatsApp) gives it significant control over information flow and communication globally. The company creates rules about data sharing, privacy, and community standards, effectively functioning as a digital nation-state. Corporate reports are available for Alphabet[43], Amazon[44], Apple[45], Broadcom[46], Meta Platforms (Facebook), Microsoft[47], Netflix[48], NVIDIA[49], TSMC[50]

The legal challenges Big Tech companies face are diverse, encompassing issues such as antitrust, data privacy, intellectual property, and content moderation. Historically and currently, various legal cases have played a significant role in shaping the operations of Big Tech companies. Below is an overview of some of the most noteworthy cases involving companies like Google, Facebook (now Meta), Apple, Amazon, and Microsoft. One of the earliest and most significant antitrust cases in the tech world was the United States v. Microsoft Corp. (1998). In this case, the U.S. Department of Justice accused Microsoft of maintaining a monopoly in the personal computer market by bundling its Internet Explorer web browser with its Windows operating system, thereby suppressing competition from other browser makers like Netscape. Microsoft was found guilty of antitrust violations. Although the company avoided being split up after an appeal, the case established essential precedents for legally challenging monopolistic practices in the tech industry.[51]

---

[43] "Investor Updates - Alphabet Investor Relations." n.d. Alphabet Investor Relations. https://abc.xyz/investor/.

[44] "Amazon.com, Inc. - Annual Reports, Proxies and Shareholder Letters." n.d. https://ir.aboutamazon.com/annual-reports-proxies-and-shareholder-letters/default.aspx.

[45] "Investor Relations - Apple." n.d. https://investor.apple.com/investor-relations/default.aspx.

[46] "Investor Center | Broadcom Inc." n.d. Broadcom Inc. https://investors.broadcom.com/.

[47] Microsoft. n.d. "Microsoft Investor Relations - Annual Reports." https://www.microsoft.com/en-us/investor/annual-reports.

[48] "Netflix - Overview - Profile." n.d. https://ir.netflix.net/ir-overview/profile/default.aspx.

[49] "NVIDIA Corporation - Financial Reports." n.d. https://investor.nvidia.com/financial-info/financial-reports/default.aspx.

[50] frontpage. n.d. "Investors - Taiwan Semiconductor Manufacturing Company Limited." Taiwan Semiconductor Manufacturing Company Limited. https://investor.tsmc.com/english.

[51] Betts, Michael. "United States versus Microsoft: A Case Study." Oklahoma Journal of Law and Technology 3, no. 1 (2007): 7.

European Union antitrust cases against Google (2010 - Ongoing) where Google has faced multiple antitrust challenges in the European Union. In 2017, the European Commission fined Google billions of dollars for favoring its comparison-shopping service in search results.[52] Facebook data privacy scandals and Cambridge Analytica in 2018, where Facebook faced legal scrutiny and a $5 billion fine from the Federal Trade Commission (FTC) for allowing data harvesting by Cambridge Analytica without user consent.[53] Apple vs. Epic Games is an antitrust case in 2020 where Epic Games challenged Apple's App Store policies, leading to a ruling requiring Apple to allow alternative payment methods. The case continues to shape discussions around app store economics and competition.[54]  Amazon faced scrutiny for promoting its products at the expense of third-party sellers. In 2023, the Federal Trade Commission filed an antitrust lawsuit against Amazon.[55]

### 3. Literature Review

#### 3.1 Existing Research: Corporate Power, Digital Sovereignty, Global Influence and Tech Companies

Europe has strongly advocated digital sovereignty, aiming to reduce its dependence on foreign technology companies and regain control over its digital landscape. The European Union (EU) is implementing regulatory frameworks to assert digital sovereignty and mitigate the dominance of big tech companies in Europe. This is part of a broader movement to reassert control over the digital economy and, importantly, to protect and uphold European values in the digital realm.[56] The EU is strategically pursuing digital sovereignty through new regulations and technological capabilities. Its approach to shaping the digital economy is designed to establish rules that limit the influence of tech giants and foster European-led innovations.[57]

---

[52] Bania, Konstantina. "The European Commissions decision in Google Search." In Competition Law for the Digital Economy, pp. 264-301. Edward Elgar Publishing, 2019.
[53] BBC News. 2019. "Facebook 'to Be Fined $5bn Over Cambridge Analytica Scandal.'" July 13, 2019. https://www.bbc.com/news/world-us-canada-48972327.
[54] United States District Court, Northern District of California. 2021. "Epic Games, Inc. V. Apple Inc. | United States District Court, Northern District of California." September 10, 2021. https://cand.uscourts.gov/cases-e-filing/cases-of-interest/epic-games-inc-v-apple-inc/.
[55] "FTC Sues Amazon for Illegally Maintaining Monopoly Power." 2023. Federal Trade Commission. December 5, 2023. https://www.ftc.gov/news-events/news/press-releases/2023/09/ftc-sues-amazon-illegally-maintaining-monopoly-power.

The growing influence of big tech companies has led to a new understanding of sovereignty, especially in the context of data and platforms. This has given way to the concept of "platform power," wherein companies like Facebook, Google, and Amazon control large digital platforms that operate as transnational entities. This concentration of power has significant implications for global governance, as these platforms wield considerable control over global data flows and economic activity[58]. Additionally, there lies in the broader impact of corporate forces on shaping digital networks and governance. As corporations gain influence over the structure and management of the internet, the state's traditional role is being challenged unprecedentedly.[59] The profound impact of corporate power, particularly in the tech industry, which has transcended national boundaries and is now shaping global governance structures. The emergence of digital sovereignty movements, notably in Europe, and the relentless expansion of tech giants' platform power are critical indicators of the shifting equilibrium between corporate influence and state sovereignty.

[56] Scott, Mark, and Konstantinos Komaitis. 2024. "Reenvisioning Europe's Digital Sovereignty." POLITICO, September 23, 2024. https://www.politico.eu/article/europe-ursula-von-der-leyen-tech-brussels-digital/.

[57] Burwell, Frances G. Digital Sovereignty in Practice: The EU's Push to Shape the New Global Economy. Atlantic Council, 2022.

[58] Gu, Hongfei. 2023. "Data, Big Tech, and the New Concept of Sovereignty." Journal of Chinese Political Science, May. https://doi.org/10.1007/s11366-023-09855-1.

[59] Sassen, Saskia. "The Impact of the Internet on Sovereignty: Real and Unfounded Worries." San Francisco: Nautilus Institute (1999).

### 3.2 Conceptual Framework: "Digital Sovereignty," "Corporate Governance," and "Platform Power"

The digital age has influenced and fundamentally transformed how governance, sovereignty, and corporate control are understood. There is an increasing focus on how platforms influence global and national policies, a shift of utmost relevance in our digitally interconnected world. This conceptual framework integrates the three interrelated concepts of digital sovereignty, corporate governance, and platform power, examining how these dynamics interact. Digital sovereignty, the ability of states or entities to control their digital infrastructure, data, and internet policies without external interference, is critical in this framework. It reflects a country's or organization's right to self-determination in cyberspace, where it can regulate digital activities within its borders and safeguard its citizens' data from foreign exploitation. The rise of multinational technology corporations, such as Google, Amazon, and Facebook, has introduced a challenge to this notion, as these platforms operate globally, often beyond the control of individual states[60]. Digital sovereignty encompasses technical control over data and governance over the platforms that collect, store, and process this data. As nations seek to enhance their digital sovereignty, they implement regulations such as data localization laws and cybersecurity frameworks, which aim to protect national interests in the face of foreign platform dominance.[61] The tension between digital sovereignty and global platforms is particularly evident in regions like the European Union, where the General Data Protection Regulation (GDPR) is an example of a legal framework designed to enhance the control of member states over personal data.[62] However, this approach to digital sovereignty raises questions about the balance between free-flowing global information and the desire for national control, as well as the practicalities of enforcing such regulations on powerful transnational platforms. The GDPR is not perfect by any stretch of the imagination, as evidenced by its complex complaints that have led to backlogs with the regulators.[63]

---

[60] Timmers, Paul and Centre on Regulation in Europe (CERRE). 2022. "Digital Industrial Policy for Europe." https://cerre.eu/wp-content/uploads/2022/12/Digital-Industrial-Policy-for-Europe.pdf.

[61] Floridi, Luciano. "The fight for digital sovereignty: What it is, and why it matters, especially for the EU." *Philosophy & technology* 33 (2020): 369-378.

Corporate governance, the system by which corporations are directed and controlled, takes on new dimensions in the context of digital platforms. These platform companies are not just economic actors but also gatekeepers of critical digital infrastructure. The governance structures of these companies determine how decisions about data privacy, content moderation, and algorithmic transparency are made, all of which impact users' rights and the broader public good. This role as gatekeepers underscores their influence on digital governance, as they have the power to shape the digital landscape and influence policy decisions. Effective corporate governance in platform companies involves balancing profitability with ethical responsibilities to users and society. However, the concentration of power within a few dominant platform companies, such as Facebook and Google, has led to concerns about the accountability of these corporations. Corporate governance failures in platforms, such as data breaches or unethical content curation, often have far-reaching consequences, including the erosion of trust and the violation of user privacy rights. Governance frameworks that promote transparency, accountability, and stakeholder engagement are crucial for ensuring that platform power is exercised in ways that align with the public interest.

---

[62] Radu, Roxana. *Negotiating internet governance*. Oxford University Press, 2019.
[63] Burgess, Matt. 2022. "How GDPR Is Failing." WIRED, May 23, 2022. https://www.wired.com/story/gdpr-2022/.

Platform power is the influence and control digital platforms exert over markets, societies, and individuals. Platforms like Amazon, Google, and Facebook have become central to modern economic and social life, serving as intermediaries that connect users to services, goods, and information.[64] This position of intermediary power allows platforms to dictate the terms of interaction, often in ways that prioritize corporate interests over public welfare.[65] Platform power is usually exercised through controlling user data, algorithms, and network effects, which enable these corporations to consolidate their market dominance and shape user behavior.[66] This concentration of platform power challenges traditional regulatory frameworks and raises significant concerns about market competition, user autonomy, and democratic governance.[67] Platforms shape economic outcomes and influence political processes, such as manipulating social media during elections or using algorithms to amplify certain types of content.[68] The power of platforms to act as arbiters of online discourse and information gatekeepers creates a new kind of governance in which private corporations play a critical role in shaping public policy and social norms[69]. The relationship between digital sovereignty, corporate governance, and platform power is deeply intertwined. As digital platforms expand their reach across borders, they increasingly challenge the sovereignty of states to control their digital ecosystems. Governments seek to regain control through regulations and laws, yet the global nature of these platforms complicates enforcement and governance. At the same time, the internal governance structures of platforms, which include decision-making processes, leadership structures, and corporate culture, determine how these corporations respond to external pressures, whether from governments, consumers, or civil society.[70] The consolidation of platform power raises fundamental questions about accountability and democratic control in the digital age. While corporate governance frameworks offer mechanisms to hold platforms accountable to their shareholders, they often fail to address platform power's broader social implications.[71] As platforms continue accumulating power, there is an increasing need for governance models that protect shareholder interests and safeguard public welfare and democratic values. Effective management in the digital age requires the integration of state-level digital sovereignty initiatives, robust corporate governance practices, and mechanisms for curbing excessive platform power, highlighting the need for a multi-layered approach to understanding the dynamics of digital sovereignty, corporate governance, and platform power. These mechanisms could include antitrust regulations, data privacy laws, and transparency requirements. By implementing such measures, we can ensure that the interests of all stakeholders, governments, corporations, and citizens—are adequately represented and protected in the digital governance landscape.

---

[64] Gillespie, Tarleton. *Custodians of the Internet: Platforms, content moderation, and the hidden decisions that shape social media*. Yale University Press, 2018.

[65] Van Dijck, José, Thomas Poell, and Martijn De Waal. The platform society: Public values in a connective world. Oxford university press, 2018.

[66] Srnicek, Nick. "Platform capitalism." Polity (2017).

[67] Kenney, Martin, and John Zysman. "The platform economy: restructuring the space of capitalist accumulation." Cambridge journal of regions, economy and society 13, no. 1 (2020): 55-76.

[68] Napoli, Philip. *Social media and the public interest: Media regulation in the disinformation age*. Columbia university press, 2019.

[69] Vaidhyanathan, Siva. Antisocial media: How Facebook disconnects us and undermines democracy. Oxford University Press, 2018.

---

[70] Zuboff, Shoshana. "The age of surveillance capitalism." In Social theory re-wired, pp. 203-213. Routledge, 2023.
[71] Taplin, Jonathan. *Move fast and break things: How Facebook, Google, and Amazon have cornered culture and what it means for all of us*. Pan Macmillan, 2017.

**3.3 Theories of Statehood and Governance about Big Tech**

The evolving relationship between governments and large technology companies, known as Big Tech, has led to various theories on statehood and governance. These theories focus on the intersection of state power, regulatory authority, and democratic principles with the increasing influence of major technology corporations such as Google, Amazon, Facebook (Meta), Apple, and Microsoft. The impact of these corporations has given rise to questioning traditional governance models and developing new frameworks to explain and address this power shift. Technological sovereignty suggests that nation-states must regain control over their technological infrastructure. Big Tech corporations, due to their global reach, have assumed quasi-sovereign roles by controlling critical aspects of the digital ecosystem, including data flows, communication channels, and economic transactions. This challenges the traditional notion of statehood as governments struggle to assert sovereignty in a domain increasingly dominated by transnational tech companies. The discussion surrounding statehood and governance in the context of Big Tech raises important questions about the future of sovereignty in the digital age. As governments grapple with how to regulate these influential entities, a complex dynamic emerges where Big Tech not only operates within national borders but also wields influence that extends beyond them. The challenge for governments is to strike a balance between fostering innovation and ensuring that these companies do not undermine the democratic principles that form the basis of statehood. The changing landscape of digital governance will likely necessitate new forms of collaboration between states, international organizations, and private corporations to address the unprecedented power dynamics brought about by Big Tech.

**3.4 Gaps in Current Research**

Studying why large companies are becoming similar to nation-states poses several challenges. The first obstacle is the inherent complexity and interdisciplinary nature of the topic. Understanding this phenomenon requires insights from economics, political science, sociology, and law, demanding a breadth of knowledge that spans multiple academic fields. Data availability and reliability also pose significant challenges. Companies often withhold full disclosure of their operations, and the available data can be incomplete or biased. The constantly shifting business and political landscapes further complicate capturing a stable and accurate picture over time.  Differences in regulations and jurisdictions across countries further complicate this research. Each nation has its legal and regulatory frameworks, making it challenging to understand these corporations' global influence and reach uniformly. Corporate secrecy also hinders research efforts. Companies are typically very private about lobbying efforts, tax strategies, and international operations, restricting the depth of any analysis. Bias and conflicts of interest can affect the research, as funding sources or entities might influence the findings.  Ethical and normative questions arise when considering the implications of corporations acting like nation-states. These questions can complicate the research process and influence interpretations and conclusions. Public opinion and political sentiment add another layer of complexity, often creating a polarized environment that affects how data is interpreted and shared. Given the limitations of accessing internal corporate information, which is critical for a comprehensive understanding, researchers must adopt a critical and multifaceted approach. This involves utilizing diverse data sources and methodologies to tackle the complex issue of corporations functioning as quasi-nation-states.

## 4. Theoretical Framework

## 4.1 State-like Characteristics: Sovereignty, Territory, Population, and Governance

The concept of nation-states as independent entities has evolved in today's global landscape, especially with the emergence of powerful multinational technology corporations. Companies like Google, Amazon, Apple, and Meta have gone beyond their corporate roles and now exhibit characteristics typically associated with nation-states. To thoroughly analyze this phenomenon, we need to consider the critical aspects of sovereignty, territory, population, and governance about how these companies embody state-like qualities. The world's largest technology companies demonstrate many of the defining characteristics of nation-states: sovereignty, territory, population, and governance. Although they may not have formal recognition, their ability to influence large populations, control digital territories, and establish governance systems positions them as pseudo-states in the global order. As their impact expands, these corporations increasingly operate as global entities with state-like powers, challenging traditional ideas of sovereignty and governance.

Sovereignty, traditionally defined as the supreme authority within a specific territory, is a fundamental characteristic of nation-states. It grants them the right to exercise power without external interference.[72] Although multinational tech giants do not formally possess legal sovereignty, their economic, informational, and technological dominance grants them de facto sovereignty in many domains. For example, these corporations operate with relative autonomy from national governments, shaping global markets, privacy norms, and political discourse. Meta, as one illustration, has implemented its community standards and policies, exerting global control over content and communications for billions of users, resembling a sovereign entity.[73] Some major tech companies now have significant influence over global regulations. For example, Apple and Google's dominance of app store ecosystems gives them substantial power over the international distribution of digital products and services. This "platform sovereignty" enables them to enforce policies and rules similar to the legislative power of nation-states.[74] Therefore, although they don't have legal sovereignty, tech giants wield power comparable to state actors, affecting international regulatory environments and consumer behavior.

[72] Krasner, Stephen D. "Sovereignty: Organized Hypocrisy." (1999).

Traditionally, territory refers to the geographical area where a state's authority exists. For a nation-state, territorial boundaries define the scope of governance and control. However, technology companies challenge the conventional notion of territoriality by creating virtual territories that transcend physical boundaries. These digital spaces, such as Amazon Web Services (AWS) or Google's cloud infrastructure, function as realms of control where the corporations govern access, usage, and data.[75] Despite not claiming physical land, these companies' digital territories are vast and pervasive, extending across national borders and controlling massive data centers globally. They often negotiate with governments to influence local laws and regulations in their favor. Their control over these digital landscapes further reinforces their state-like characteristics, as they wield significant influence over global internet infrastructure, giving them control over a territory or territories not constrained by geography in a traditional sense but resemble a traditional nation-state or states.[76]

---

[73] Bradshaw, Samantha, and Philip N. Howard. "The global disinformation order: 2019 global inventory of organised social media manipulation." (2019).

[74] Zuboff, Shoshana. "The age of surveillance capitalism." In Social theory re-wired, pp. 203-213. Routledge, 2023.

[75] Srnicek, N. "Platform Capitalism. Polity Press, Cambridge Malden, MA." (2017).

A vital aspect of a nation-state is its population, which serves as its social and political foundation. Similarly, tech companies' user bases are their populations, encompassing billions of individuals. For example, Meta has several billion active users, a population more significant than any single country on Earth.[77] These users are not just passive consumers; they actively engage with the platform, generate data, and participate in the digital ecosystem, much like citizens engage in their country's civic and economic life. These companies' influence over their user populations is extensive, impacting consumer behavior and even democratic processes, as evident in the manipulation of political ads and the spread of misinformation on social media platforms.[78] In essence, tech giants have established global communities that resemble the populations of nation-states, exerting power through data management, service provision, and communication facilitation.

[76] Plantin, Jean-Christophe, Carl Lagoze, Paul N. Edwards, and Christian Sandvig. "Infrastructure studies meet platform studies in the age of Google and Facebook." New media & society 20, no. 1 (2018): 293-310.

[77] Statista. 2024. "Facebook: Quarterly Number of MAU (Monthly Active Users) Worldwide 2008-2023." May 21, 2024. https://www.statista.com/statistics/264810/number-of-monthly-active-facebook-users-worldwide/.

[78] Vaidhyanathan, S. (2018). Antisocial Media: How Facebook Disconnects Us and Undermines Democracy. Oxford University Press.

**4.2 Figure 1. Number of monthly active Facebook users worldwide as of 4th quarter 2023 (in millions)**

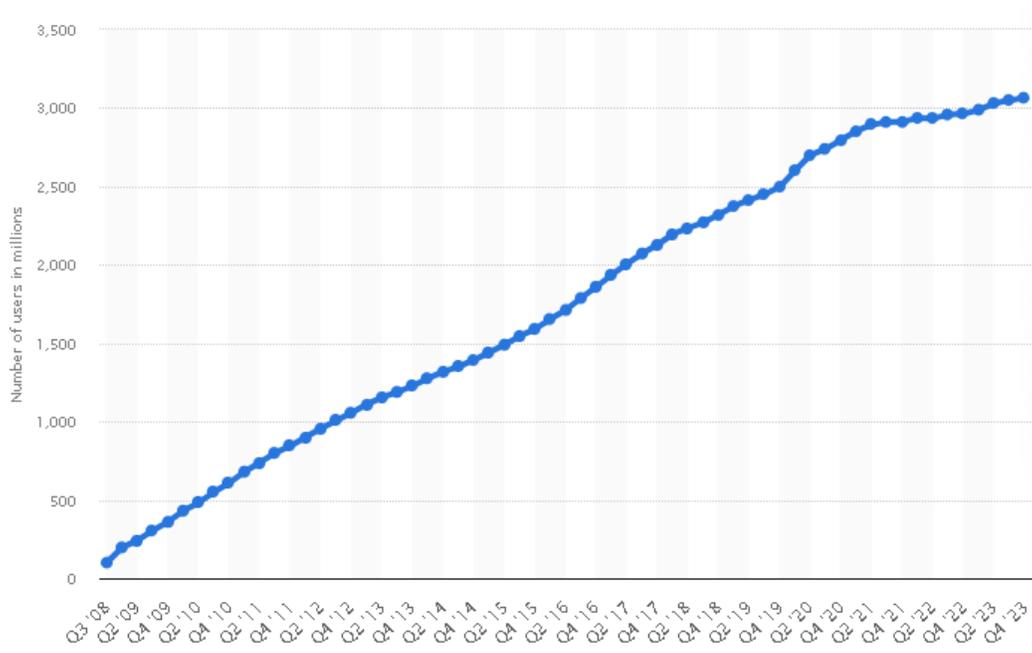

79

Governance is the process through which a state exercises authority and upholds laws. Nation-states govern their populations through political systems, legal frameworks, and institutions. Similarly, tech companies have developed complex governance systems to regulate platform behavior. These systems encompass the terms of service, privacy policies, algorithms, and AI-driven moderation tools. They determine what is allowed, how conflicts are resolved, and how information flows within the digital environments they oversee.[80]  For instance, social media platforms like YouTube and Facebook have content moderation policies. Human moderators and algorithms interpret and enforce these policies, deciding what content is acceptable and what violates platform rules. In this way, tech companies act like governments, creating and enforcing regulations that impact billions of lives.[81] These platforms sometimes have quasi-judicial bodies, such as Facebook's Oversight Board, which reviews and overturns content decisions—a notable parallel to nation-state judicial institutions.[82]

[79] Statista. 2024. "Facebook: Quarterly Number of MAU (Monthly Active Users) Worldwide 2008-2023." May 21, 2024. https://www.statista.com/statistics/264810/number-of-monthly-active-facebook-users-worldwide/.

[80] Gillespie, Tarleton. Custodians of the Internet: Platforms, content moderation, and the hidden decisions that shape social media. Yale University Press, 2018.

### 4.3 Global Influence of Tech Companies: Economic, Social, and Political Dimensions

Information dissemination, organization, and narrative control have undergone profound transformations through platforms like Meta's Facebook, X (formerly Twitter), and YouTube. Social media has played a vital role in mobilizing and coordinating movements, amplifying marginalized voices, and internationalizing conflicts. This includes the spread of misinformation and propaganda via recommendation engines, which has sometimes contributed to creating and prolonging disputes. [83] Large corporations' actions can have significant legal and moral implications, such as concerns over privacy, labor rights, environmental impact, and the influence of recommendation engines that can influence people as propaganda. These issues often transcend national borders, requiring international cooperation and regulation.

The Chinese-owned video-sharing app TikTok censors politically sensitive content from the Chinese government. The app's moderation guidelines restrict or ban material related to the Tiananmen Square protests, Tibetan independence, and the religious group Falun Gong, as well as claims of censorship of protests in Hong Kong for political reasons. Bytedance is the parent company of TikTok and denies such claims. The company acknowledged that in the early days of TikTok, there was a broad approach to minimizing conflict on the platform, including penalties for content promoting conflict between religious sects or ethnic groups. However, Bytedance stated that they have since recognized the need for a more nuanced and localized approach. The app, launched in China in 2016 as Douyin, has grown rapidly and now has half a billion active users, 40% outside China. Despite the app's popularity, concerns about its content moderation policies have been raised, particularly regarding political sensitivity.[84] Nonetheless, there remains hope for empowering grassroots movements and exposing injustices, highlighting the dual-edged nature of these platforms' immense power. In either case, their profound influence remains undeniable. Giant corporations lobby, shape regulatory environments, and influence tax laws and trade agreements, which mirrors nation-states' diplomatic and policy maneuvers.[85] Multinational corporations, operating in numerous countries and often having a presence on every continent, have a global reach that allows them to impact economies, labor markets, and cultures worldwide, like the influence exerted by powerful nations.[86] These large corporations frequently control vast resources, including natural resources, intellectual property, and human capital, and their management of these resources can significantly affect global supply chains, technological advancements, and market dynamics. [87] Tech giants have developed global technologies integral to everyday life, managing massive amounts of data, controlling critical infrastructure like cloud services, and leading innovation in AI, communication, and other fields.[88] Many corporations and organizations operate with a level of autonomy and self-regulation that mirrors the governance structures of sovereign states. They establish their policies, codes of conduct, and ethical guidelines, often enforcing them independently of national laws.[89] The concept of companies as nation-states underscores their significant role and impact on the global stage, challenging traditional notions of sovereignty and power.[90] In many Arab Spring countries, the government tightly controlled traditional media outlets, limiting public access to information and providing a narrow, state-sanctioned narrative of events. Social media, however, allowed citizens to bypass these restrictions. In countries like Tunisia, Egypt, and Libya, activists used Facebook and Twitter to document events on the ground and counteract the official narratives pushed by state-run news agencies.[91]

[81] Tufekci, Zeynep. Twitter and tear gas: The power and fragility of networked protest. Yale University Press, 2017.
[82] Gorwa, Robert, Reuben Binns, and Christian Katzenbach. "Algorithmic content moderation: Technical and political challenges in the automation of platform governance." Big Data & Society 7, no. 1 (2020): 2053951719897945.
[83] Bradshaw, Samantha, and Philip N. Howard. "The global organization of social media disinformation campaigns." *Journal of International Affairs* 71, no. 1.5 (2018): 23-32.

---

[84] Hern, Alex. 2019. "Revealed: How TikTok Censors Videos That Do Not Please Beijing." The Guardian, September 25, 2019. https://www.theguardian.com/technology/2019/sep/25/revealed-how-tiktok-censors-videos-that-do-not-please-beijing.

[85] Coen, David, and Alexander Katsaitis. "Between cheap talk and epistocracy: The logic of interest group access in the European Parliament's committee hearings." Public Administration 97, no. 4 (2019): 754-769.

[86] Scherer, Andreas Georg, and Guido Palazzo. "The new political role of business in a globalized world: A review of a new perspective on CSR and its implications for the firm, governance, and democracy." Journal of management studies 48, no. 4 (2011): 899-931.

These digital ecosystems grant them significant influence over national and international policies, elections, and, at times, civil wars and revolutions. The Arab Spring, which unfolded across the Middle East and North Africa between 2010 and 2012,[92] represents one of the most significant waves of social and political unrest in the early 21st century. This series of uprisings was characterized by widespread protests calling for democratic reforms, overthrowing authoritarian regimes, and greater social justice. Central to the success and momentum of these movements was the use of digital platforms and technologies, mainly social media. The role of big tech companies and social media platforms such as Facebook, Twitter, and YouTube in shaping, facilitating, and amplifying the Arab Spring is a critical area of study for understanding how digital activism operates in authoritarian contexts. This thesis explores how social media and big tech companies empowered and sometimes constrained revolutionary movements during the Arab Spring.  The ability of social media to amplify voices beyond national borders created a unique global dynamic, where information flowed out of authoritarian states and into the hands of foreign audiences. This internationalization of the protests also encouraged solidarity movements in other parts of the world, highlighting the global dimensions of digital activism of creating change for the better or chaos for the worse.

---

[87] Sassen, Saskia. "The many scales of the global: Implications for theory and for politics." Critical globalization studies (2005): 155-66.
[88] Zuboff, Shoshana. "Surveillance capitalism and the challenge of collective action." In New labor forum, vol. 28, no. 1, pp. 10-29. Sage CA: Los Angeles, CA: SAGE Publications, 2019.
[89] Cutler, A. Claire. Private power and global authority: Transnational merchant law in the global political economy. Vol. 90. Cambridge University Press, 2003.
[90] Vogel, David. "The private regulation of global corporate conduct." (2006).
[91] Howard, Philip N., and Muzammil M. Hussain. "The upheavals in Egypt and Tunisia: The role of digital media." Journal of democracy 22, no. 3 (2011): 35-48.
[92] Borg Cardona, Karen. "The Failed Quest in Contemporary World Literature." https://core.ac.uk/download/266992442.pdf.

### 4.4 Parallels Between Tech Companies and Traditional Nation-States

These big tech companies' global reach is evident in their operations across numerous countries, detailed in annual reports and international trade data.[93]  Major tech companies in the United States lobby the federal government on issues that affect their business. In 2023, Meta was the internet industry's largest lobbyist spender in the US, investing $19.3 million in lobbying efforts during the calendar year. Amazon.com was ranked second with $19.27 million U.S. dollars in lobby spending, followed by Alphabet at $12.35 million. This helps provide perspective on the complex interplay between corporate entities and national governments and

---

[93] Gordon, Ian. "Porter, ME (1990), The Competitive Advantage of Nations, Macmillan."

the broader implications for global governance and economic stability. Big Tech companies such as Google, Meta, and Amazon have been accused of using deceptive lobbying practices in the European Parliament. Lawmakers have filed complaints against eight companies and trade groups, alleging they engaged in shadow lobbying. The accusations claim that Big Tech companies lobbied through smaller front organizations, misleading lawmakers and deceiving them during negotiations on two significant EU tech laws, the Digital Markets Act (DMA) and Digital Services Act (DSA). The companies allegedly concealed their involvement in these lobbying efforts by funding and instructing smaller lobbying groups representing small and medium-sized companies.[94] It's safe to assume with moderate confidence that money can be spent anywhere globally, influencing its needs, and big tech is spending there.

Much like a state's intelligence apparatus, tech companies exercise vast surveillance powers through their ability to collect, store, and analyze personal data. The big tech companies have access to unprecedented information about individual behavior, preferences, and social interactions, allowing them to profile citizens precisely so that surveillance can monetize data to predict and shape human behavior.[95] The level of data control enjoyed by these companies can rival or exceed the capacity of traditional governments. With insights derived from personal data, tech firms can shape advertising, market preferences, and even political campaigns, impacting democratic processes like state intelligence agencies intervening in foreign elections or policymaking.[96] Interestingly, in some cases, nation-states are embracing tech nationalism, supporting their domestic tech industries to assert influence on the global stage. For instance, the Chinese government's promotion of Huawei as a national champion in the tech sector illustrates how the state can use tech companies to project global power and maintain economic independence.[97] Simultaneously, Huawei was caught in a dispute between the United States and China, which restricted its business with United States companies. Google blocked Huawei's access to Android updates, and many retailers and networks stopped working with Huawei due to fear of United States sanctions.[98] The United States has revoked export licenses for Intel and Qualcomm to supply Huawei with semiconductors, increasing pressure on the Chinese company. This affects Huawei's supply of chips for its laptops and mobile phones amid concerns about Huawei's chip development capabilities in the United States. The Chinese commerce ministry opposes the United States' move, describing it as economic coercion.[99]

[94] Goujard, Clothilde. 2022. "Big Tech Accused of Shady Lobbying in EU Parliament." POLITICO, October 17, 2022. https://www.politico.eu/article/big-tech-companies-face-potential-eu-lobbying-ban/.

[95] Zuboff, Shoshana. "The age of surveillance capitalism." In Social theory re-wired,. Routledge, 2023.

[96] Cadwalladr, Carole, and Emma Graham-Harrison. "Revealed: 50 million Facebook profiles harvested for Cambridge Analytica in major data breach." The Guardian 17, no. 1 (2018): 22.

---

[97] Feldstein, Steven. The rise of digital repression: How technology is reshaping power, politics, and resistance. Oxford University Press, 2021.

[98] Beavis, Gareth, and Basil Kronfli. 2019. "Huawei Ban: The Global Fallout Explained." TechRadar, October 2, 2019. https://www.techradar.com/news/huawei-ban.

[99] "US Revokes Licences for Supply of Chips to China's Huawei." May 8, 2024. Financial Times. https://www.ft.com/content/cf965960-b083-49ee-bae1-6ce95fe872a3.

**4.5.1 Figure 2, Lobbying expenses of the leading internet companies in the United States in 2023 (in million U.S. dollars)**

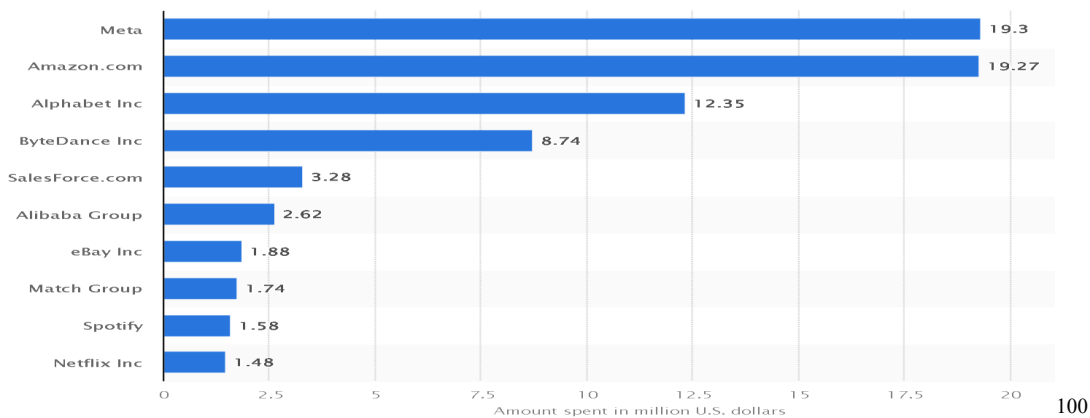

Corporate governance structures, outlined in publicly available reports, show their self-regulatory practices. Legal cases and regulatory actions also provide evidence of these companies' ethical and legal challenges.[101] These various sources substantiate the view that large corporations function similarly to nation-states regarding economic influence, global reach, and governance. Additionally, many of these companies engage in substantial philanthropy, claim to invest in renewable energy, and lead in corporate responsibility efforts.[102] Nation-states maintain embassies and negotiate international treaties, but tech companies also engage in diplomatic activities. Major tech corporations have expanded their reach to global markets, offering services that span borders and impact lives in countries where they may not have a physical presence.

---

[100] Statista. 2024. "Top Internet Lobbying Spenders in the U.S. 2023." May 24, 2024. https://www.statista.com/statistics/1035987/us-leading-internet-lobbying-spenders/.
[101] Brammer, Stephen, Gregory Jackson, and Dirk Matten. "Corporate social responsibility and institutional theory: New perspectives on private governance." Socio-economic review 10, no. 1 (2012): 3-28.
[102] Carroll, Archie B. "Corporate social responsibility: The centerpiece of competing and complementary frameworks." Organizational dynamics (2015).

These companies often interact with international governments and regulators, negotiating trade deals, resolving disputes, and navigating complex geopolitical environments similar to sovereign states. For example, Facebook has been involved in discussions with multiple governments regarding data privacy, misinformation, and the influence of its platform on national elections. Many tech companies have launched initiatives in emerging markets, providing internet access in remote areas (e.g., Elon Musk's Starlink, the satellite internet service), effectively shaping infrastructure like nation-state foreign aid programs.

While tech companies do not control physical territory like nation-states, they dominate digital spaces and increasingly essential technological infrastructures. Google, for example, controls over 90% of global search engine activity, acting as a gatekeeper to the information people access online. Amazon Web Services (AWS) powers many cloud infrastructures that businesses and governments depend on, while Apple maintains tightly controlled hardware and software ecosystems. These companies' control over essential digital infrastructures has been likened to the control nation-states exert over critical physical resources such as oil or land.[103] Nation-states typically build on a shared national identity, often defined by language, culture, and history[104]. Big tech companies, on the other hand, promote a global digital culture that transcends national identities. Platforms like YouTube, Instagram, and TikTok foster communities where people from different nationalities interact with similar content, participate in global trends and form new identities  less tied to traditional national boundaries.  The concept of the nation-state, defined by its sovereignty, shared national identity, centralized governance, and nationalism, is being tested by the growing influence of large tech companies that operate globally and prioritize their corporate interests over national priorities. As these companies expand their control over global communication, data, and economic systems, they challenge traditional state sovereignty, governance, and identity ideas, creating complex power dynamics between governments and tech corporations. In this new digital era, nation-states must find ways to reaffirm their authority while managing the impact of powerful multinational corporations.

---

[103] Zuboff, Shoshana. "The Age of Surveillance Capitalism: The Fight for a Human Future at the New Frontier of Power, edn." PublicAffairs, New York (2019).
[104] Gellner, Ernest. Nations and nationalism. Cornell University Press, 2008.

## 5. Historical Development of Nation-States and Technology Companies

### 5.1 History of Nation-States: The Evolution of Sovereignty and Governance

The concept of the nation-state, as understood today, is a relatively recent development in the long history of human governance. Its evolution has been marked by profound transformations in political, social, and economic structures, particularly around sovereignty and governance. The nation-state is characterized by a defined territorial boundary, a centralized government, and a sense of national identity shared by its citizens. This model emerged from the complex interplay of historical forces in Europe during the early modern period and gradually spread across the globe. Before the rise of the nation-state, medieval Europe was dominated by a patchwork of feudal systems, where political power was fragmented among local lords, religious authorities, and monarchs. Sovereignty during this period was not concentrated in a single entity but shared among different levels of authority. For example, kings often partnered with the Church and local nobles who controlled specific territories. Political loyalty was primarily personal, with subjects pledging allegiance to individual rulers rather than abstract political entities.[105] The transition toward centralized states began in the late Middle Ages and the early modern period, particularly with the consolidation of royal power in countries like France and England. The growth of taxation systems, standing armies, and bureaucratic institutions enabled monarchs to exert greater control over their territories, reducing the influence of local nobility and the Church. This era witnessed the strengthening of sovereign rule and the gradual formation of distinct political entities that resemble modern nation-states.[106]

A pivotal moment in the nation or nation-state's evolution came with the Peace of Westphalia in 1648, which ended Europe's Thirty Years' War. This treaty is often cited as the foundation of the modern international system of states. The Peace of Westphalia introduced the principle of state sovereignty, recognizing the right of rulers to govern their territories free from external interference. This marked a significant shift in the understanding of governance, as states were now considered the primary actors in international relations, with clear territorial boundaries and exclusive authority over their domestic affairs.[107] The Westphalian model also established the notion of legal equality among states, regardless of their size or power, and the norm of non-intervention in the internal affairs of other states. These sovereignty and territorial integrity principles became central to the international system, forming the basis for developing the modern nation-state.[108]

---

[105] Tilly, Charles. "Coercion, capital, and European states, AD 990–1990." In Collective violence, contentious politics, and social change, pp. 140-154. Routledge, 2017.
[106] Anderson, Benedict. "Imagined communities: Reflections on the origin and spread of nationalism." In The new social theory reader, pp. 282-288. Routledge, 2020.
[107] Osiander, Andreas. "Sovereignty, international relations, and the Westphalian myth." International organization 55, no. 2 (2001): 251-287.
[108] Krasner, Stephen D. "Sovereignty: Organized Hypocrisy." (1999).

The next major shift in the history of the nation-state occurred with the rise of nationalism in the 18th and 19th centuries. Nationalism is the belief that people with a common language, culture, and history should have their state. This idea gained momentum during that time, known as the Enlightenment era when legitimate government must be based on the will of the people rather than divine right or hereditary privilege[109]. In particular, the French Revolution of 1789 was a turning point in this regard, as it popularized the idea of the nation as a community of citizens bound together by shared rights and duties. The revolutionaries declared that sovereignty resided in the people, not the monarch, and this principle spread across Europe, inspiring nationalist movements in countries like Italy, Germany, and Greece. The 19th century saw the unification of Italy and Germany, which were previously fragmented into smaller states and principalities, further cementing the idea of the nation-state as the dominant form of political organization.[110]

The nation-state model expanded globally during the 19th and 20th centuries, mainly through decolonization after World War II. As European empires collapsed, former African, Asian, and Middle East colonies sought to establish independent nation-states. These newly formed states adopted the Westphalian principles of sovereignty and territorial integrity, even as many struggled to build cohesive national identities despite ethnic, linguistic, and religious diversity.[111] The United Nations, founded in 1945, further enshrined the nation-state as the central actor in global politics. The UN Charter recognized the sovereignty of all its member states and emphasized the importance of territorial integrity and non-interference, reinforcing the international legal framework established by the Peace of Westphalia. However, the postcolonial period also highlighted the challenges of nation-state formation, as many newly independent states faced internal conflicts and pressures. In the late 20th and early 21st centuries, the idea of the nation-state has been questioned due to globalization. Economic integration, the growth of multinational corporations, and supranational organizations such as the European Union have all undermined the traditional limits of sovereignty. Some experts argue that we are moving towards a post-Westphalian world, where the power of nation-states is increasingly dispersed and challenged by non-state actors and global institutions.[112]

[109] Smith, Anthony D. "National Identity University of Nevada Press." Nevada, USA (1991).
[110] Breuilly, John, ed. The Oxford handbook of the history of nationalism. OUP Oxford, 2013.
[111] Jackson, Robert H. Quasi-states: sovereignty, international relations and the Third World. Vol. 12. Cambridge university press, 1990.

## 5.2 The Rise of Tech Companies: From Startups to Global Empires

The rise of tech companies from small startups to global empires is a defining narrative of the 20th and 21st centuries, marked by innovation, disruption, and an unprecedented transformation of international markets. It began with a few ambitious entrepreneurs operating out of garages or a backyard shed, as in the case of Hewlett-Packard (HP). Eventually, it created some of the world's most valuable companies, reshaping industries and societies. The origins of modern tech giants can be traced back to the 1970s, primarily in Silicon Valley, California, which emerged as the global epicenter of technological innovation. This region's risk-taking, entrepreneurship, and venture capital culture fueled the early tech boom. Companies like Intel and HP were among the first to establish a stronghold, laying the groundwork for a new era of technology.[113] The 1970s also saw the rise of Apple Inc., founded by Steve Jobs and Steve

---

[112] Held, David. Democracy and the global order: From the modern state to cosmopolitan governance. Stanford University Press, 1995.
[113] Mazzucato, Mariana. "The entrepreneurial state." Soundings 49, no. 49 (2011)

Wozniak in 1976 with its original help from Xerox. Apple's launch of the Apple II in 1977 marked one of the first successful mass-produced personal computers, along with the IBM PC, Atari, Commodore, and Radio Shack TSR-80 computers, which began revolutionizing how people interacted with technology.[114] Around the same time, Microsoft was founded by Bill Gates and Paul Allen in 1975, and it became the dominant software provider with its MS-DOS and later Windows operating systems. MS-DOS was initially paired with the IBM PC, but many companies were able to produce hardware running the same operating system. The late 1990s brought the dot-com boom, fueled by the rapid expansion of the internet. Companies like Amazon, founded by Jeff Bezos in 1994, and Google, launched by Larry Page and Sergey Brin in 1998, began as startups with novel ideas about e-commerce and search engine technology, respectively. Amazon started as an online bookstore, while Google revolutionized internet search by developing algorithms that vastly improved search results.[115]

This period was characterized by immense speculation, leading to the dot-com bubble—a speculative stock market bubble in internet-related companies. Many companies failed when the bubble burst in 2000, but survivors like Amazon and Google thrived, learning from the era's mistakes. By the early 2000s, both companies had begun diversifying their product offerings, which set the stage for them to grow into global conglomerates. The rise of social media and mobile technology in the mid-2000s catalyzed the next wave of tech giants. Facebook, founded by Mark Zuckerberg in 2004, quickly became the world's largest social networking platform, connecting billions of people and redefining how society communicates. Twitter/X in 2006 and Instagram in 2010 followed, contributing to the social media boom and creating a new digital ecosystem.[116] This resilience of tech companies during the dot-com bubble reassures us about the stability and adaptability of the tech industry, even in the face of significant challenges. In parallel, Apple's introduction of the iPhone in 2007 sparked a revolution in mobile technology. The smartphone became the primary computing device for billions, creating new industries, such as mobile apps, and allowing companies like Facebook and Google to extend their influence through mobile platforms. Android, developed by Google, became the dominant mobile operating system, propelling Google into a central role in the mobile economy.[117]

[114] Dyck, Jeremy. 2022. "The Xerox Thieves: Steve Jobs & Bill Gates - BC Digest - Medium." Medium, March 30, 2022. https://medium.com/bc-digest/the-xerox-thieves-steve-jobs-bill-gates-6e1b36fc1ec4.
[115] Perry, Alex. 2019. "What Google, Amazon, and Apple Were Doing 20 Years Ago." Mashable, August 30, 2019. https://mashable.com/article/apple-amazon-google-1999-then-now?test_uuid=01iI2GpryXngy77uIpA3Y4B&test_variant=a.

By the 2010s, the largest tech companies—now known collectively as Big Tech (Google, Amazon, Facebook, Apple, and Microsoft, often referred to as GAFA (Google, Amazon, Facebook, Apple or FAANG- N for Netflix)—began dominating the tech industry and the entire global economy. These companies expanded into cloud computing, artificial intelligence (AI), and big data. Amazon Web Services (AWS), Google Cloud, and Microsoft Azure are now critical infrastructure providers for businesses worldwide.[118] The 2010s also saw NVIDIA transition from gaming GPUs to AI hardware. GPUs like the Tesla series became crucial for large-scale scientific computations, deep learning, and AI workloads in data centers. NVIDIA's GPUs played a significant role in the modern AI boom, enabling advances in autonomous vehicles, natural language processing, medical research, and other AI-driven fields. This advancement has seen NVIDIA's growth skyrocket to a current market capitalization of over three trillion US dollars. [119] In addition to cloud dominance, artificial intelligence has become a focal point for tech innovation. Google, for example, acquired DeepMind in 2015, making it a leader in AI research. Similarly, Amazon's use of AI in its supply chain and Apple's development of AI for Siri and other applications showcase how these companies leverage cutting-edge technology to maintain a competitive edge.[120] As these companies grow, their influence extends beyond commerce and technology into social, political, and cultural realms, prompting debates over their role in modern society. Their journey from scrappy startups to global empires reflects broader trends in globalization, innovation, and the shifting balance of economic power. As we move further into the 21st century, the trajectory of tech companies continues to evolve. The integration of artificial intelligence into everyday life and the increasing importance of data privacy and security are just a few of the ongoing developments shaping the tech industry's future.[121]

---

[116] Boyd, Danah M., and Nicole B. Ellison. "Social network sites: Definition, history, and scholarship." Journal of computer-mediated Communication 13, no. 1 (2007): 210-230.
[117] Isaacson, Walter. Walter Isaacson Great Innovators e-book boxed set: Steve Jobs, Benjamin Franklin, Einstein. Simon and Schuster, 2011.
[118] Srnicek, Nick. "Platform capitalism." Polity (2017).

### 5.3 Parallels Between Historical State Formation and Corporate Growth

Digital sovereignty has become crucial in international relations, significantly impacting global governance and diplomacy. Unlike traditional sovereignty, which involves clearly defined authority based on geographical boundaries, digital sovereignty is more complex and often overlapping. National governments face the challenge of enforcing local laws in a global digital environment where data flows do not respect physical borders. As a result, new forms of cooperation and conflict arise as states attempt to exert control over their digital territories. Initiatives like data localization, where countries require specific types of data to be stored domestically, illustrate the tension between national interests and the inherently global nature of digital networks.[122] The evolution toward digital sovereignty indicates a shift in the role of nation-states as they navigate the delicate balance between openness and control in a globalized digital economy. While nation-states once primarily concerned themselves with physical borders, they are now increasingly preoccupied with digital boundaries. This adaptation acknowledges that data and digital infrastructure are vital to national security and economic competitiveness. As digital technology advances, the notion of sovereignty will likely continue to evolve, underscoring the persistent role of nation-states in shaping, regulating, and securing the digital domains within their influence.

[119] Leswing, Kif. 2024. "Nvidia Closes at Record as AI Chipmaker's Market Cap Tops $3.4 Trillion." CNBC. October 14, 2024. https://www.cnbc.com/2024/10/14/nvidia-shares-hit-a-record-as-chipmaker-market-cap-tops-3point4-trillion.html.
[120] Goodfellow, Ian. "Deep learning." MIT Press, (2016).
[121] Cassidy, John. Dot. con: How America lost its mind and money in the internet era. HarperCollins Publishers, 2003.

Digital territory refers to online spaces mainly controlled by big tech companies instead of traditional governments. Unlike physical land with clear borders, these digital areas are global. They include data centers, social media platforms, and software systems. These spaces organize, control, and give access to personal data and digital services, typically managed by private companies. This situation creates new challenges related to authority and control, often leading to conflicts between tech companies and national governments regarding data privacy, security, and regulations.

---

[122] Chander, Anupam, and Uyên P. Lê. "Data nationalism." Emory LJ 64 (2014): 677.

Companies like Google, Meta (formerly Facebook), Amazon, and Apple gather large amounts of user data to improve their services, target advertising, and enhance user experiences. They set policies and algorithms that govern data collection, user access, and privacy. By doing this, tech giants control digital areas and greatly influence users. This power often leads to tensions with local governments, which want more oversight and regulatory power, resulting in data privacy, security, and authority disputes.

Big tech companies create rules, guidelines, and terms of use that dictate how users behave on their platforms. For example, Twitter (now X), Facebook, and Instagram set standards for acceptable content. This gives these companies significant power to regulate that can go beyond the reach of national governments. These companies build and manage data centers worldwide, forming the physical foundation of digital territory. Platforms like Amazon Web Services, Microsoft Azure, and Google Cloud illustrate this system, allowing data to be stored, processed, and accessed across borders, creating a borderless digital world. The algorithms used by big tech companies strongly impact the content people see online.[123] These algorithms shape public opinion and can influence political outcomes by promoting some content while filtering out others. This informational governance acts like how historical empires expanded their influence. It can significantly affect society.[124]

The parallels between nation-states' history and big tech companies' rise are striking. Both entities assert control, exercise power, and navigate challenges in governance and sovereignty. As nation-states historically held territorial sovereignty, Big Tech companies have created their forms of sovereignty in the digital sphere. Both have evolved in response to shifts in power dynamics, technological advances, and the pressures of globalization. In the early development of nation-states, centralization of power was vital to establishing sovereignty and governance. Similarly, Big Tech companies like Google, Amazon, Facebook (Meta), and Apple consolidated their influence through strategic acquisitions, massive data networks, and control over critical infrastructure. Big Tech companies exert power through the collection of user data and control over digital platforms, much like how early modern states used taxation and standing armies to extend their control. Nation-states and Big Tech companies negotiate privacy, antitrust, and content regulation issues with governments and international bodies. The rise of nationalism in the 19th century bears similarities to the strong brand loyalty seen among Big Tech users. Both have fostered a sense of belonging and identity among their users, sometimes called brand tribalism.[125] Big Tech companies have extended their influence globally, usually superseding the power of national governments. For example, Facebook wields immense power over communication and information dissemination worldwide.[126]

[123] Bilić, Paško. "Search algorithms, hidden labour and information control." Big Data & Society 3, no. 1 (2016): 2053951716652159.
[124] Zuboff, Shoshana. "The age of surveillance capitalism." In Social theory re-wired, pp. 203-213. Routledge, 2023.

## 6. Economic Power and Market Sovereignty

### 6.1 Global Economic Influence: Tech Giants vs. Nation-States' GDP Contributions today and in history

The most obvious parallel between large tech companies and nation-states is their vast economic power. Apple, for example, reached a market capitalization of over $3 trillion in 2022,

---

[125] Francis, Abey. 2024. "Brand Tribalism &Ndash; Consumer Tribal Behavior on Brand Loyalty." MBA Knowledge Base, June. https://www.mbaknol.com/marketing-management/brand-tribalism/.

[126] Flew, Terry, Fiona Martin, and Nicolas Suzor. "Internet regulation as media policy: Rethinking the question of digital communication platform governance." Journal of Digital Media & Policy 10, no. 1 (2019): 33-50.

making it wealthier than the GDP of many mid-sized countries, including Spain and Australia. Similarly, Amazon and Microsoft control global infrastructures—Amazon with its dominance in e-commerce and cloud services and Microsoft with its enterprise software solutions. This economic control gives these companies leverage in markets and politics, as they can lobby for favorable regulations or exemptions, much like how powerful nations advocate for their interests on the global stage.[127]

The top ten countries' GDP (in trillions) in today's values are the United States $28.78, China $18.53, Germany $4.59, Japan $4.11, India $3.94, United Kingdom $3.50, France $3.13, Brazil $2.33, Italy $2.33 and Canada $2.24. [128] The top seven of the eight top companies by market capitalization are tech companies at today's values (in trillions): Apple $3.529, NVIDIA $3.358, Microsoft $3.097, Alphabet $2.014, Amazon $1.968, Meta Platforms $1.459, and TSMC $1.067. All except TSMC (from Tawain) are from the United States.[129] The Figures below show a graphical representation of a world map's Gross Domestic Product from 2021, with Apple, Microsoft, Amazon, and Facebook as nation-states. The pictures speak a thousand words about how big these companies have become compared to and perhaps as nation-states. The companies are on par with the United Kingdom and France in terms of size monetarily.

[127] Srnicek, N. "Platform Capitalism. Polity Press, Cambridge Malden, MA." (2017).
[128] Silver, Caleb. 2024. "The Top 25 Economies in the World." Investopedia. October 5, 2024. https://www.investopedia.com/insights/worlds-top-economies/.
[129] "Companies Ranked by Market Cap - CompaniesMarketCap.com." n.d. https://companiesmarketcap.com/.

### 6.1.1 Figure 3, Visualizing historic big companies versus today's Big Tech companies.[130]

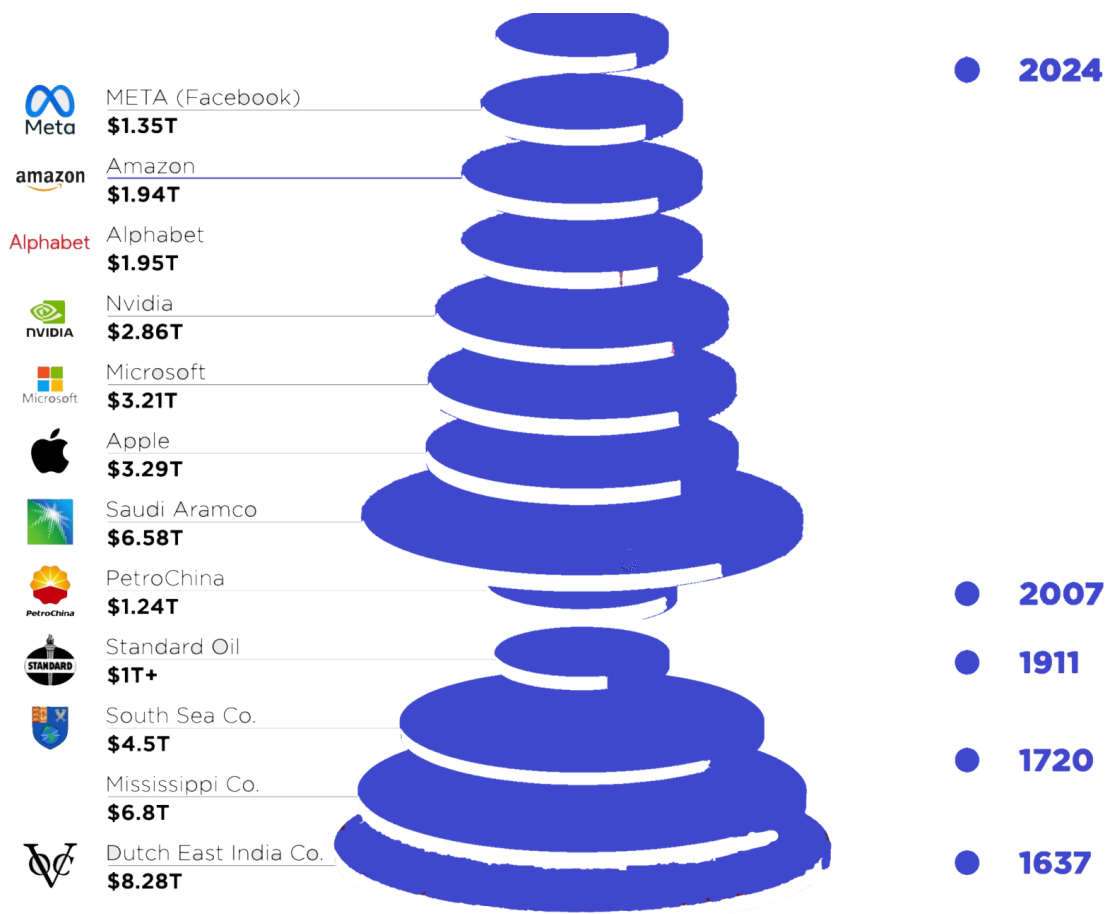

Historically, Big Tech boasts some of the biggest companies ever. Only a few are bigger historically; the Dutch East India Company, the Mississippi Company, and the South Sea Company were historical financial giants whose market valuations, at their peak, were the only companies historically that surpassed those of today's largest corporations. The similarities

---

[130] "Visualizing Top 20 Most Valuable Companies of All Time." 2024. HowMuch. October 10, 2024. https://howmuch.net/articles/the-worlds-biggest-companies-in-history.

between these early enterprises and today's big tech firms are their immense market power, speculative bubbles, and monopolistic tendencies that dominated their respective eras. The comparisons between these entities across time provide critical insights into the workings of capital markets, speculative investments, and the pursuit of monopolistic control. The Dutch East India Company and the South Sea Company held monopolies granted by their respective governments. The Dutch East India Company was founded in 1602, became the world's first publicly traded company, and had exclusive rights to trade in Asia. This control over lucrative trade routes allowed it to amass enormous wealth and influence in global commerce, equivalent to about $8 trillion in today's terms at its peak. Similarly, today's tech companies like Google, Apple, Amazon, and Facebook exert monopoly-like control over critical industries through e-commerce, digital advertising, or cloud computing services—paralleling how the Dutch East India Company dominated spice and silk trade.[131] The Mississippi and South Sea Companies are infamous for their role in two of history's most significant financial bubbles. Both companies promised immense profits based on speculative ventures in the New World (Mississippi Co.) or debt restructuring (South Sea Co.). Fueled by greed and the promise of extraordinary returns, investors poured money into these companies, driving their valuations to unsustainable levels before their inevitable collapse in 1720. Similarly, the dot-com bubble of the late 1990s and early 2000s mirrored this phenomenon, as tech startups saw inflated valuations based on speculative promises of future growth, only to collapse when those expectations were not met. Although today's tech companies are more grounded in profitable business models, the soaring valuations of companies like Tesla or the widespread adoption of cryptocurrencies evoke echoes of past

---

[131] Petram, Lodewijk. The world's first stock exchange. Columbia University Press, 2014.

speculative manias[132]. Both early companies and modern tech giants were seen as innovators that could open up new markets. The Dutch East India Company pioneered global trade networks, while the Mississippi and South Sea Companies promised to unlock the riches of untapped lands. In the 21st century, tech companies have revolutionized industries, from artificial intelligence to social media, creating entirely new economic landscapes. The optimism surrounding these new technologies has led to market exuberance that is reminiscent of the enthusiasm investors had for global trade and exploration in the 17th and 18th centuries. Just as governments were deeply involved in the establishment and support of the Dutch East India Company, Mississippi Company, and South Sea Company, today's tech companies benefit from favorable regulation and, in some cases, direct government support. For instance, the Dutch government granted the VOC military protection and legal monopolies, while modern companies like Amazon benefit from lucrative government contracts. This intertwining of state and corporate interests helped both sets of companies cement their dominance and expand beyond the constraints of purely private enterprise. In conclusion, these early behemoths shared with modern tech giants a reliance on monopoly power, speculative investment, the promise of innovation, and government backing. While modern companies operate in a more regulated and complex global financial system, the fundamental dynamics of market dominance and speculative excess continue to draw striking parallels with the past.

---

[132] Garber, Peter M. Famous first bubbles: The fundamentals of early manias. mit Press, 2001.

**6.1.2 Figure 4, Illustration of the market cap of the biggest tech companies compared to the world's GDP.**

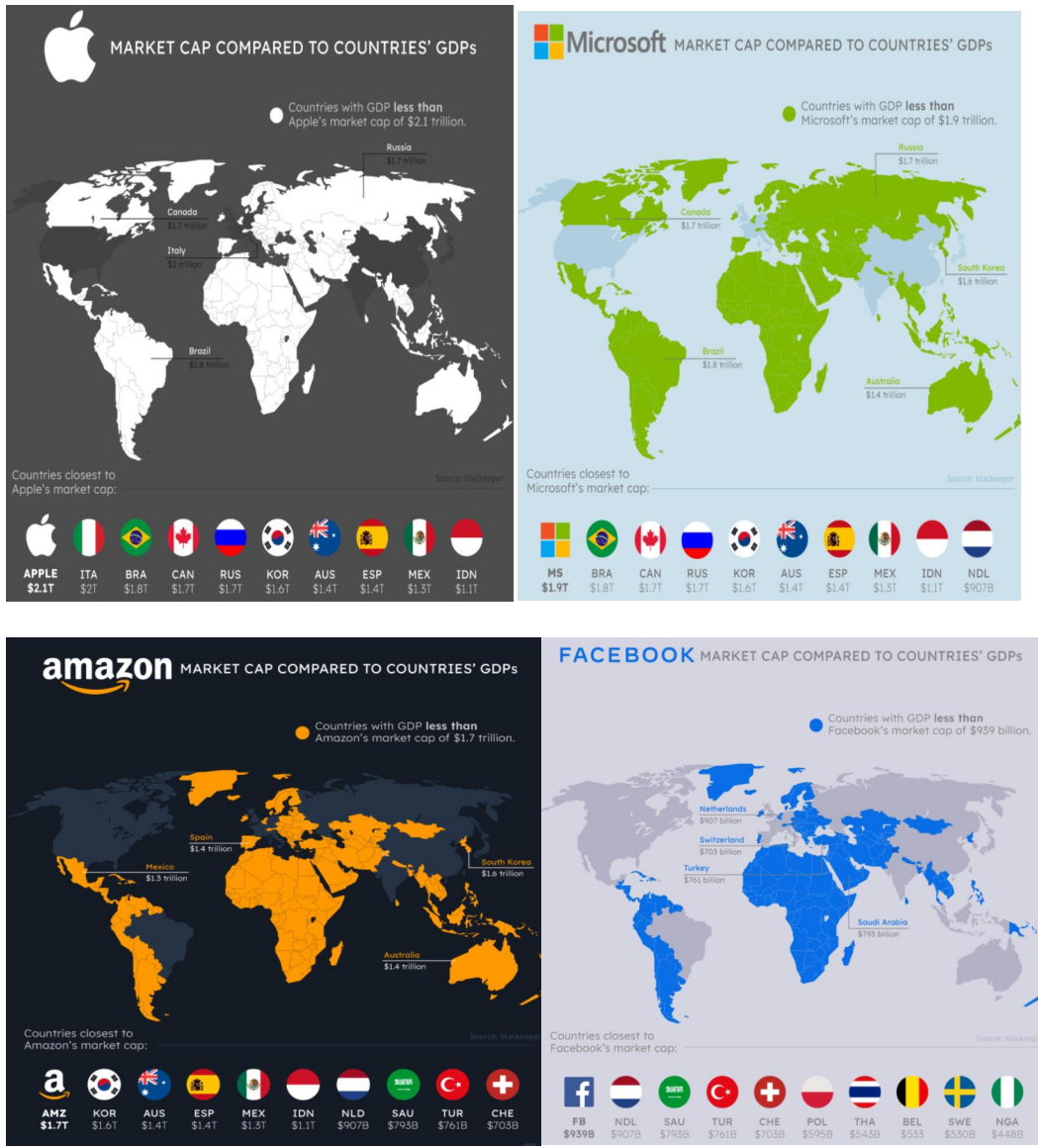

These current technology companies rival historical levels for being the most valuable companies ever. The only three companies that surpass Apple, Microsoft, NVIDIA, Alphabet,

133 Wallach, Omri. 2021. "The World's Tech Giants, Compared to the Size of Economies." Visual Capitalist. July 9, 2021. https://www.visualcapitalist.com/the-tech-giants-worth-compared-economies-countries/.

and Amazon in order adjusted to today's market capitalization in trillions are The Dutch East India Company, $8.28, The Mississippi Company, $6.58, and The South Sea Company, $4.5 all historical notorious monopolies.The only current company that hovers near the top tech companies is Saudi Aramco, which has its roots in Standard Oil. [134]

### 6.2 Monopoly and Market Control: Digital Market Domination

These corporations have grown to wield an influence and economic power that rivals or surpasses some countries, leading to significant parallels in how they operate, govern, and influence global affairs. Large tech companies and nation-states' most apparent similarity is their immense economic power. This financial control gives these companies leverage in markets and politics, allowing them to lobby for favorable regulations or exemptions, much like how powerful nations advocate for their interests on the global stage.[135]  Like countries like the U.S. or China project cultural influence globally, tech companies wield immense cultural power. Their platforms and services shape how people communicate, consume media, and think about themselves and their identities. Facebook's and Instagram's algorithms influence what news people see and how they interact with others, such as how state-run media channels operate in authoritarian countries.  As of 2024, there are approximately 5.16 billion active social media users worldwide, around 59.3% of the global population. The average daily time spent on social media was 2 hours and 23 minutes worldwide. People aged 16-24 spent the most time on social media, at an average of 3 hours and 38 minutes daily. Most social media is from the top tech giants, with Meta (Facebook and Instagram) and Google (YouTube), followed by TikTok and others. Meta's Facebook leads the pack with 3.04 billion users, maintaining its position as the most extensive social networking site globally. Alphabet/Google's YouTube follows with 2.5 billion users, reinforcing its status as the premier platform for video sharing and consumption. WhatsApp and Meta's Instagram are tied for third place, each with 2 billion users. WhatsApp is renowned for its messaging services, while Instagram is a favorite for photo and video sharing. With 1.5 billion users, TikTok rounds out the top five, showcasing its rapid rise as a leading platform for short-form video content. The growth of this field is showing projections to double in ten years from 2017 to 2027.[136] An ongoing antitrust case involving Meta's acquisitions of WhatsApp and Instagram has raised questions about whether Meta is a monopolist in the personal social networking services market. The FTC alleges that Meta has monopoly power, but Meta argues that it does not meet the criteria for monopolization. The FTC's case relies on defining a narrow market for services that includes only a few platforms. At the same time, Meta contends that it competes with many other social media firms in various areas. The outcome of this case could have significant implications for interpreting antitrust laws and regulating Big Tech companies.[137]

---

[134] "Visualizing Top 20 Most Valuable Companies of All Time." 2024. HowMuch. October 10, 2024.
https://howmuch.net/articles/the-worlds-biggest-companies-in-history.
[135] Srnicek, N. "Platform Capitalism. Polity Press, Cambridge Malden, MA." (2017).

---

[136] Larson, Stefan. 2024. "Social Media Users 2024 (Global Data & Statistics)." Priori Data. April 8, 2024. https://prioridata.com/data/social-media-usage/.

[137] Coniglio, Joseph V. 2024. "Is Meta Really a Monopoly? Debunking the FTC's Market Definition Metaphysics." ITIF. July 6, 2024. https://itif.org/publications/2024/06/06/is-meta-monopoly-debunking-ftc-market-definition-metaphysics/.

**6.2.1 Figure 5, Social Media Users from 2017 projected through 2027**

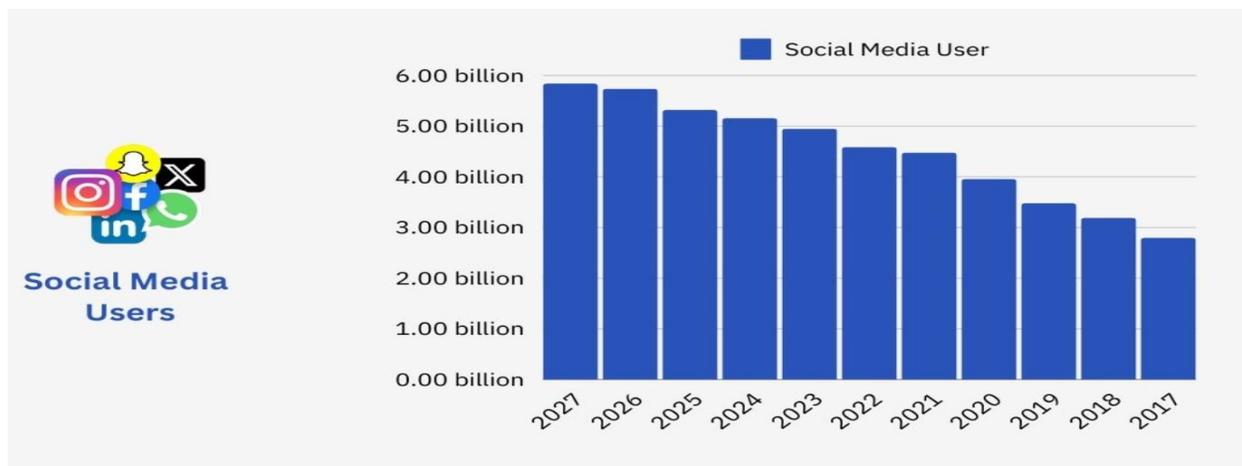

The global popularity of platforms like Meta (Facebook, What's App, Instagram), YouTube, TikTok, and X/Twitter are part of the broader ecosystem of digital platforms that contribute to spreading ideas and culture akin to the soft power wielded by nation-states.[138] Several countries have banned TikTok due to concerns about privacy, security, and the spread of inappropriate content. Afghanistan, Australia, Belgium, Canada, Denmark, European Union, France, India, Indonesia, Latvia, Netherlands, Nepal, New Zealand, Norway, Pakistan, Somalia, and Taiwan have all taken measures such as banning TikTok from government-issued devices or imposing nationwide bans. These actions were taken in response to worries about cybersecurity, privacy, misinformation, and the promotion of immoral or offensive content.[139] The rise of the most significant technological companies has prompted comparisons between these corporate giants and traditional nation-states. These corporations have reached a scale of influence and economic power that rivals or surpasses some countries, leading to significant parallels in how they operate, govern, and influence global affairs.

[138] Nye, Joseph S. Soft power: The means to success in world politics. Public affairs, 2004.
[139] "These Countries Have Already Banned TikTok." 2024. PBS News. April 26, 2024.
https://www.pbs.org/newshour/world/these-countries-have-already-banned-tiktok.

Tech giants also mimic nation-states through their governance structures. As governments create rules to regulate behavior within their borders, tech companies develop and enforce policies governing user behavior. Facebook, for instance, created its own "Supreme Court," known as the Oversight Board, to handle disputes about content moderation. The company imposes rules on what content is acceptable on its platforms, akin to how a government legislates speech, all while influencing the lives of billions of users. In this way, tech companies wield quasi-sovereign authority over digital spaces.[140] While tech companies do not control physical territory like nation-states, they dominate digital spaces and increasingly essential technological infrastructures. Google, for example, controls over 90% of global search engine activity, acting as a gatekeeper to the information people access online. Amazon Web Services (AWS) powers many cloud infrastructure businesses and governments depend on, while Apple maintains a tightly controlled hardware and software ecosystem. These companies' control over essential digital infrastructures has been likened to the control nation-states exert over critical physical resources such as oil or land.[141]

Tech giants often negotiate with governments and international organizations and employ lobbyists, lawyers, and diplomats to influence policy, particularly around issues like taxation, antitrust law, and data privacy. In some cases, tech companies have more leverage than smaller countries. For example, Facebook and Google have negotiated user data, privacy, and digital advertising regulations with governments. These corporations influence international discussions about data sovereignty and digital governance, often negotiating deals or settling fines akin to treaties.[142] Tech companies increasingly provide essential public goods and services. Google Maps and Microsoft's Azure cloud services are integrated into the functioning of many public and private institutions. During the COVID-19 pandemic, e-commerce and Amazon's logistics network were crucial for supplying essential goods, while tech platforms were controversial and important in public communication and information dissemination. These services have become so ubiquitous that society depends on them for basic functionality, which parallels how populations rely on governments for public services[143].

---

[140] Klonick, Kate. "The new governors: The people, rules, and processes governing online speech." Harv. L. Rev. 131 (2017): 1598.
[141] Zuboff, Shoshana. "The Age of Surveillance Capitalism: The Fight for a Human Future at the New Frontier of Power, edn." PublicAffairs, New York (2019).

### 6.3 Taxation and Revenue: Navigating International Tax Laws Like Nation-States

Tech giants navigate tax obligations like certain countries seeking to lower their tax burdens through offshore accounts and intricate tax policies; companies have been criticized for using legal loopholes and offshore tax havens to minimize tax payments. This has led to tensions with nation-states, especially within the European Union, where there have been multiple high-profile cases of tech companies being fined for tax avoidance. The taxation disputes echo nation-states' challenges in enforcing tax laws on global enterprises.[144]

---

[142] Moore, Martin, and Damian Tambini, eds. Digital dominance: the power of Google, Amazon, Facebook, and Apple. Oxford University Press, 2018.
[143] Crawford, Kate. The Atlas of AI: Power, Politics, and the Planetary Costs of Artificial Intelligence. Yale University Press, 2021.
[144] Picciotto, Sol. International business taxation. London: Weidenfeld & Nicolson, 1992.

**6.4 Case studies revealing anti-competitive behavior of Big Tech**

The Standing Committee on Finance in India submitted its report on 'Anti-Competitive Practices by Big Tech Companies' in December 2022. The committee has identified ten types of anti-competitive practices by Big Tech companies. App Stores such as Google and Apple App Store publishers prevent their business users (app users) from leaving the platform and using alternative payment methods, known as anti-steering behavior. Commonly associated with digital marketplaces, some platforms prominently place their products. For example, Google placed Google Pay prominently on the Play Store. Digital firms force people to buy related services. For instance, food delivery apps require restaurants to use the platform's services—monopolistic data usage by digital firms, especially from leading platforms with vast data repositories.  Big Tech buys highly valued start-ups, disallowing smaller firms to grow beyond a specific limit, such as Facebook's acquisition of WhatsApp. E-commerce sales as one example, Amazon has huge discounts that compromise service providers' control over the final price and offline players' ability to compete. An exclusive arrangement of e-commerce platforms with a brand hampers the business of other platforms and brick-and-mortar sellers. Similarly, platforms use price parity clauses to stop businesses from selling at lower rates on different platforms. Via search platforms, preference is given to sponsored products in algorithms used to show results for users' search rather than organic search results. For example, Amazon and any Google search starting with sponsored searches. The practice of restricting third-party applications whereby digital markets restrict the installation or operation of third-party applications. For example, Apple's App Store is the only channel for app developers to distribute their apps to iOS consumers. When platforms operate at all levels of the ad-tech supply chain, the digital advertising market faces a conflict of interest and self-preferencing issues.

India is working on regulating big tech and safeguarding competition through strengthening competition laws and safe harbor provisions for the digital era. In the United States, antitrust legislation targets the dominance of Big Tech companies. This legislation gives states greater power in competition cases and increases funding for federal regulators. In Australia, the competition watchdog in Australia has recommended tighter regulation of Facebook and Google and efforts to improve media competition. Additionally, the Online Safety Act will have the power to compel social media companies to delete posts that amount to online bullying and to hold the companies and those who hosted the alleged abuse accountable. In Europe, the Digital Markets Act (DMA) will ban harmful business practices by very large digital players and create a fairer and more competitive economic space for new players and European businesses. The Digital Services Act targets categories of online services, from simple websites to internet infrastructure services and online platforms.[145]

## 7. Conclusion

### 7.1 Summary of Findings: Big Tech's Function as Nation-State Equivalents

We can choose between allowing powerful corporations to pursue their interests or shaping markets to ensure that technology benefits the public and is controlled democratically by citizens rather than corporations. Competition policy is the most effective tool for reshaping markets in this manner. It's important to note that this doesn't just involve antitrust laws and

---

[145] PWOnlyIAS. 2024. "Big Tech: Impact, Dominance and Anti-Competitive Practices - PWOnlyIAS." PWOnlyIAS, July 24, 2024. https://pwonlyias.com/big-tech/.

regulations but requires a comprehensive whole-of-government approach to address the various threats posed by the dominance of these large technology companies. This approach should encompass powers related to privacy, consumer protection, corporate governance, copyright law,

trade policy, labor law, and industrial policy. By closely integrating these systems in the United

States and elsewhere, we can more quickly and effectively ensure that technology serves the interests of the people rather than just the interests of the largest corporations.[146]  A comprehensive policy framework must encompass various areas of governance beyond just competition law. Privacy regulation is essential, especially as tech companies amass vast amounts of personal data that can be exploited to undermine individual autonomy and privacy[147].

Consumer protection is equally critical, as large corporations often manipulate or obscure

information to the detriment of users' ability to make informed choices.[148] Corporate governance

reforms must be part of this strategy to ensure that technology companies operate with

transparency and accountability, balancing their profit motives with public interests.[149]

Intellectual property laws, including copyright regulations, must be modernized to prevent a few dominant entities from monopolizing knowledge and culture. Reforming copyright law can ensure a more equitable distribution of the benefits of innovation while fostering creative expression and fair competition. [150] Trade policies must be aligned with these goals, addressing the global nature of tech monopolies and ensuring that international trade agreements do not undermine domestic regulatory efforts[151].  Labor law also plays a crucial role in addressing how automation, AI, and other technologies impact workers. Industrial policy can and should support the development of alternative technologies and business models that prioritize public welfare, sustainability, and fair labor practices.[152]

[146] Markets, Open. 2024. "Report | AI in the Public Interest: Confronting the Monopoly Threat — Open Markets Institute." Open Markets Institute. April 25, 2024. https://www.openmarketsinstitute.org/publications/report-ai-in-the-public-interest-confronting-the-monopoly-threat.

[147] Pasquale, Frank. The black box society: The secret algorithms that control money and information. Harvard University Press, 2015.

[148] Wu, Tim. The curse of bigness: Antitrust in the new gilded age. Vol. 15. New York: Columbia Global Reports, 2018.

[149] Bruner, Christopher M. n.d. "Corporate Governance Reform and the Sustainability Imperative." https://www.yalelawjournal.org/feature/corporate-governance-reform-and-the-sustainability-imperative.

[150] ElJurdi, Fadi. n.d. "Copyright in the Age of Artificial Intelligence." https://jurdilaw.com/blogs/copyright-in-the-age-of-artificial-intelligence.html.
[151] Stiglitz, Joseph E. Globalization and its discontents revisited: Anti-globalization in the era of Trump. WW Norton & Company, 2017.
[152] Nissim, Gadi, and Tomer Simon. 2021. "The Future of Labor Unions in the Age of Automation and at the Dawn of AI." Technology in Society 67 (September): 101732. https://doi.org/10.1016/j.techsoc.2021.101732.